\documentclass[a4paper,12pt]{article}
\usepackage{indentfirst}
\usepackage[margin=1.1in]{geometry}
\usepackage{graphicx} 
\usepackage{amssymb}
\usepackage{lscape}
\usepackage{amsmath}
\usepackage{multirow}
\usepackage{fancyhdr} 
\usepackage{apacite}
\usepackage{subfigure}
\usepackage{float}
\usepackage{mathrsfs}
\usepackage{algorithm}
\usepackage[noend]{algpseudocode}
\usepackage{textcomp}
\usepackage{setspace}

\begin{document}
\title{Variable Selection and Missing Data Imputation in Categorical Genomic Data Analysis by integrated 
Ridge Regression and Random Forest}
\author{Siru Wang\thanks{Siru Wang is with the School of Mathematics and Statistics,
The University of Melbourne, VIC 3010, Australia}
and Guoqi Qian\footnote{Corresponding author; email: qguoqi@unimelb.edu.au} 
\thanks{Guoqi Qian is with the School of Mathematics and Statistics,
The University of Melbourne, VIC 3010, Australia.}
}
\date{}

\maketitle

\begin{abstract}
Genomic data arising from a genome-wide association study (GWAS) are often not only of large-scale, but also
incomplete. A specific form of their incompleteness is missing values with non-ignorable missingness mechanism.
The intrinsic complications of genomic data present significant challenges in developing an unbiased and informative
procedure of phenotype-genotype association analysis by a statistical variable selection approach. In this paper
we develop a coherent procedure of categorical phenotype-genotype association analysis, in the presence of
missing values with non-ignorable missingness mechanism in GWAS data, by integrating the
state-of-the-art methods of random forest for variable selection, weighted ridge regression with EM
algorithm for missing data imputation, and linear statistical hypothesis testing for determining the missingness
mechanism. Two simulated GWAS are used to validate the performance of the proposed procedure.
The procedure is then applied to analyze a real data set from breast cancer GWAS.
\end{abstract}

\section{Introduction} \label{sec-1}
A single-nucleotide polymorphism (SNP) refers to a genetic variant at a locus on a chromosome pair which should 
be present in at least $1\%$ of a population. It is known that a genetic disease or phenotype may be intricately 
related to a small set of SNPs. A number of genome-wide association studies (GWAS) have been carried out in 
genomics for identifying such SNPs that are associated with the risk of a phenotype. However, genomics data 
collected for GWAS often contain missing values on SNPs all over the place due to limitations of DNA sequencing
techniques. Ignoring the missing data in GWAS is likely to give biased or even misleading information to the 
downstream work, especially when the missing data are not missing at random (NMAR) and the missingness 
mechanism is not ignorable.

In cases of the missing data being missing (completely) at random (MAR or MCAR), relatively simple methods are
available for accommodating them for GWAS. For example, a so-called single imputation method is often used
where each missing value of a SNP for an individual can be replaced by this SNP's most frequent value observed
in the population stratum this individual belongs to. See e.g. chapter 7 of \citeA{foulkes2009applied} and \texttt{R}
package \texttt{pan} of \citeA{grund2016multiple} for details.

In cases of handling NMAR missing values of SNPs in GWAS, one needs a statistical model to describe the 
randomness involved in the SNPs having missing values, and another model to characterize the missingness
mechanism. \citeA{baker1988regression} used a log-linear regression model for analyzing categorical variables
involving non-ignorable missing values. Joint probability distribution of the covariates having missing values is 
modelled by the product of a sequence of one-dimensional conditional distributions in \citeA{lipsitz1996conditional}.
This approach is extended by \citeA{ibrahim1999missing} to model the missingness mechanism as well as the
covariates having missing values. It was further extended by \citeA{stubbendick2003maximum} to handle the
missing values and missingness mechanism in random effect models.

Since the SNPs having missing values in a GWAS may be large in number and widespread across the
population with the missingness mechanism unclear, one should resort to statistical models to describe the
joint probability distributions of the missing values and the missingness mechanism, from which multiple
rather than single imputations can be generated to substitute the missing SNPs values involved in each individual.

Several computational tools, such as \texttt{fastPhase} \cite{scheet2006fast} and \texttt{Impute}
\cite{marchini2007new}, have been developed to impute SNP missing values based on a statistical model
for the SNPs, where the linkage disequilibrium (LD) between SNPs is also taken into account. The \texttt{Emlasso}
developed by \citeA{sabbe2013emlasso} does the imputation implicitly based on the EM algorithm 
\cite{Dempster1977EM} and a generalized location model (GLoMo) \cite{olkin1961multivariate}. However, these
imputation methods overlook the missingness mechanism that may be non-ignorable. In this paper, we propose
a new method of multiple imputations through integrating a current such method with multiple logistic models
for the missingness mechanism. Ridge regression \cite{hoerl1981ridge} will be used to fit the resultant model 
if the number of covariates involved is large.

Since we use multiple logistic models to describe the SNPs missingness mechanism, we are able to apply a
well-developed linear hypothesis testing method to test whether or not the missingness of a SNP variable depends
significantly on any available SNP variables or the phenotype variable. Therefore, a conclusion on whether or
not the missingness mechanism of a SNP variable is ignorable can be drawn based on statistical evidence. 
This is an advantage of modelling the missingness mechanism by a parametric statistical model.

The set of SNPs significantly associated with the phenotype could be inferred from applying a variable
selection procedure to the joint log-likelihood function combining the phenotype-genotype conditional
association probabilities, the joint probability of the SNPs having missing values, and the conditional
probability of the missingness mechanism. But this is computationally infeasible because the dimension of
the selection space is proportional to the number of SNPs under investigation, leading to combinatorial 
explosion in computing complexity. In this paper, we propose to use a supervised machine learning
approach, to be briefed below, for SNPs selection after we substitute the generated multiple 
imputations for the missing values. This approach is shown to be computationally feasible.

The phenotype-SNPs association can be delineated by some machine learning models, upon which SNPs 
variable selection can be computationally feasibly performed. An examples is \texttt{lasso}
\cite{tibshirani1996regression} when the phenotype-SNPs association is represented by a logistic regression 
model. Another example is \texttt{elastic-net} \cite{zou2005regularization} which is effective for variable
selection when the SNPs in the logistic model are not only of high-dimensional but also highly correlative.

Ensemble classification and regression trees (CART) model gives a more flexible description of the
phenotype-SNPs association, and is likely to result in better prediction performance. Feasible variable
selection also can be performed on an ensemble CART model. For example, in random forest (RF) model
proposed by \citeA{breiman2001random} one can compute a variable importance (\texttt{VIMP}) measure
for each SNP variable, and rank all SNPs in the model by the \texttt{VIMP} measure. Accordingly, the most
important SNPs associated with the phenotype are selected to be those top ranked.

The main ideas of RF are two-folded: (i) Bootstrapping the data sample, by which a tree is constructed for
each bootstrap sample of the data; (ii) Sub-sampling the predictors (mostly referred to the SNPs here-and-after)
so that a randomly selected subset of the SNPs are used to decide the best split at each node-splitting stage. 
The out of bag (OOB) errors are then aggregated across all trees in an RF to give a total prediction error of the 
model, from which a \texttt{VIMP} measure for each SNP is derived for variable selection. More details on 
the \texttt{VIMP} measure are given in section~\ref{sec-2.2}.

\citeA{chen2012random} reviews the applications of RF for genomic data analysis and variable selection,
where three influential RF techniques for cases of high-dimensional and highly correlative SNPs are discussed
in detail. The first one is the gene-shaving random forest (GSRF) of \citeA{jiang2004joint}. GSRF performs
iterations of RF and backward variable elimination/selection by: (i) Re-computing the retained SNPs' 
\texttt{VIMP} values after each SNPs backward elimination; (ii) Computing the total prediction errors based on 
both the OOB samples and independent test samples. The second one is the gene selection random forest
(GeneSrF) of \citeA{diaz2006gene}, which enhances the computing efficiency over GSRF by: (i) computing the
\texttt{VIMP} values in the first iteration of RF only; (ii) the prediction errors being computed on only the OOB 
samples. The third one is from \citeA{calle2011auc} which replaces the mis-classification error used in the previous
two techniques with the ROC curve's AUC as the prediction accuracy measure. We choose to use the GeneSrF
technique for its performance regarding computing efficiency.
 
In this paper, we develop a coherent procedure of categorical phenotype-SNPs association analysis, in the presence of
missing values with non-ignorable missingness mechanism in GWAS data, by integrating the
state-of-the-art methods reviewed in this section, namely, random forest for variable selection, weighted ridge
regression with EM algorithm for missing data imputation, and linear statistical hypothesis testing for determining 
the missingness mechanism.
The paper is organized as follows. We first develop the proposed method in Section~\ref{sec-2}. In Section~\ref{sec-3},
we conduct two simulation studies to assess the method's performance and illustrate its use in practice. Then we apply 
the proposed method to analyze a real data set from the Australian breast cancer and mammographic density
GWAS in Section~\ref{sec-4}. Finally, Section~\ref{sec-5} discusses the improvements and future work extending 
the proposed method. 

\section{Method} \label{sec-2}
The method we are to develop is motivated by the need of addressing the missing value problem in the Australian 
Breast Cancer Family Study (ABCFS) \cite{Dite2003} and the Australian Mammographic Density Twins and Sisters
Study (AMDTSS)\cite{Odefrey2010}. A specific data subset from ABCFS and AMDTSS  contains observations of
207 SNPs on a specific gene pathway, suspected to be susceptible to breast cancer, for 596 individuals 
comprising 354 breast cancer patients and 242 matched controls. Each SNP variable for an individual in the data
takes a value from 0, 1 and 2, representing the number of the minor alleles at the individual's SNP loci. The
response variable in the data is binary indicating whether an individual has breast cancer or not. The data
contain 1,724 missing values out of 123,372 SNP observations. The proportion of missing values is about
1.4\% but they distribute across all SNPs and individuals widely and irregularly. Specifically, 157 SNPs have
missing values with each SNP being missing in between 1 to 32 individuals; 513 individuals have missing SNP
values with each individual having 1 to 9 such missing SNP values; 50 SNPs have no missing values on all 596
individuals; and 83 individuals have no missing values on all 207 SNPs.

It is clear that we first need to develop a statistical model capable of characterizing those data as seen above,
which may involve non-ignorable missing values and large number of highly correlative SNPs, 
in order to effectively analyze them and find out which predictor variables are significantly associated with
the response variable. Section~\ref{sec-2.1} below will present this model together with its fitting procedure and our
proposed missing value multiple imputation method. In section~\ref{sec-2.2} we develop an RF induced variable
selection procedure for identifying the most important predictors to the response variable. The variable
selection and missing value multiple imputation are executed in iteration until convergence. The corresponding
computational method is also presented in section~\ref{sec-2.2}. Then we present a linear hypothesis testing procedure
in section~\ref{sec-2.3} to infer the missingness mechanism.


\subsection{Model and notations} \label{sec-2.1}

Let $y$, $\mathbf{z}$ and $\mathbf{x}$ denote a binary response variable having no missing values, a predictor
vector having no observed missing values, and a predictor vector having observed missing values, respectively.
In the context of GWAS data analysis, $y=1$ indicates the presence, and $y=0$ indicates the absence of a 
phenotype (e.g. breast cancer). In the same context, $\mathbf{z}$ and $\mathbf{x}$ refer to the SNPs in the data.
Recall that a SNP variable is coded as the number of minor allele at a locus, thus takes 3 possible values 0, 1 and 2,
representing the three states of the alleles pair: homozygous recessive, heterozygous, and homozygous dominant.  
\citeA{gauderman2007testing} used an alternative way to code a SNP to express it as either dominant or recessive.

We choose to code each SNP by two dummy variables $I(\mbox{SNP}=1)$ and $I(\mbox{SNP}=2)$. Thus, assume
that $\mathbf{z}=(1, z_1, z_2, \cdots, z_{2q-1}, z_{2q})^\top$ is a vector of dummy variables plus constant 1 
for $q$ SNPs (denoted as $\mbox{SNP}^1, \cdots, \mbox{SNP}^q$) having no missing values, and 
$\mathbf{x}=(x_1, x_2,\cdots, x_{2p-1}, x_{2p})^\top$ is a vector of dummy variables for $p$ SNPs 
(denoted as $\mbox{SNP}_1, \cdots, \mbox{SNP}_p$) having missing values. 
Then for each $j$, $\mbox{SNP}_j=0\,\Leftrightarrow\, (x_{2j-1}=x_{2j}=0)$,
$\mbox{SNP}_j=1\,\Leftrightarrow\, (x_{2j-1}=1, x_{2j}=0)$, and 
$\mbox{SNP}_j=2\,\Leftrightarrow\, (x_{2j-1}=0, x_{2j}=1)$. The same equivalences hold for each $\mbox{SNP}^j$
as well.

Denote $y_i$, $\mathbf{z}_i=(1, z_{i,1}, \cdots, z_{i,2q})^\top$ and 
$\mathbf{x}_i=(x_{i,1},\cdots, x_{i,2p})^\top$ as the $i$th observation of $y$, $\mathbf{z}$ and $\mathbf{x}$,
respectively, with individual $i=1,\cdots, n$. The $n$ observations of $(y, \mathbf{z}, \mathbf{x})$ can be 
reasonably assumed to be independently and identically distributed in a case-control study. 
If denote $\mbox{SNP}_{ji}$ as the real value of $\mbox{SNP}_j$ for individual $i$. Then we have
$\mbox{SNP}_{ji}=0\,\Leftrightarrow\, (x_{i,2j-1}=x_{i,2j}=0)$,
$\mbox{SNP}_{ji}=1\,\Leftrightarrow\, (x_{i,2j-1}=1, x_{i,2j}=0)$, and 
$\mbox{SNP}_{ji}=2\,\Leftrightarrow\, (x_{i,2j-1}=0, x_{i,2j}=1)$. The same equivalences hold for each 
$\mbox{SNP}^{ji}$ as well.
Since $\mathbf{x}$ has missing values, let $\mathbf{r}=(r_1,\cdots, r_p)^\top$ be the missingness indicator
vector of the $p$ SNPs; and $\mathbf{r}_i=(r_{i1},\cdots, r_{ip})^\top$ be the $i$th observation of $\mathbf{r}$,
i.e. $r_{ij}=1$ if the $\mbox{SNP}_j$ value of individual $i$ is missing, and $r_{ij}=0$ otherwise,
$j=1,\cdots, p$; $i=1,\cdots, n$. For convenience of presentation, we write $\mathbf{y}=(y_1,\cdots,y_n)^\top$,
$Z=(\mathbf{z}_1,\cdots, \mathbf{z}_n)^\top$, $X=(\mathbf{x}_1,\cdots, \mathbf{x}_n)^\top$ and
$R=(\mathbf{r}_1,\cdots, \mathbf{r}_n)^\top$ in the sequel.

Given $\mathbf{z}_i$ and $\mathbf{x}_i$, $y_i$ follows a Bernoulli distribution with probability of phenotype
$\pi_i=\mbox{Pr}(y_i=1|\mathbf{z}_i, \mathbf{x}_i)$ which satisfies a logistic regression model
\begin{equation}
\log\frac{\pi_i}{1-\pi_i}=(\mathbf{z}_i^\top, \mathbf{x}_i^\top)\pmb{\beta}, \quad i=1,\cdots, n.\quad
\Longleftrightarrow\quad \log\frac{\pmb{\pi}}{\mathbf{1}-\pmb{\pi}}=(Z, X)\pmb{\beta}
\label{eq-1}
\end{equation}
where $\pmb{\pi}=(\pi_1, \cdots, \pi_n)^\top$, and $\pmb{\beta}$ is a $(1+2q+2p)\times 1$ unknown 
parameter vector specifying the effects of the SNPs on the phenotype which is to be estimated from the data. 
By (\ref{eq-1}) the joint conditional probability of $\mathbf{y}$ given $(Z, X)$ is
\begin{equation}
p_1(\mathbf{y}|Z, X, \pmb{\beta})=\prod_{i=1}^n p_1(y_i|\mathbf{z}_i,\mathbf{x}_i, \pmb{\beta})
=\prod_{i=1}^n\pi^{y_i}(1-\pi_i)^{1-y_i}.
\label{eq-2}
\end{equation}
Since each $\mathbf{x}_i$ contains missing value(s), a joint probability distribution is needed to model
$\mathbf{x}_i$. Using the approach of \citeA{lipsitz1996conditional}, the joint probability of $\mathbf{x}_i$
equals the product of a sequence of conditional trinomial probabilities. Denote
\begin{eqnarray}
\phi_{ij1}&=&\mbox{Pr}(x_{i,2j-1}=1, x_{i,2j}=0|\mathbf{x}_{i,1:2(j-1)}), 
\nonumber \\
\phi_{ij2}&=&\mbox{Pr}(x_{i,2j-1}=0, x_{i,2j}=1|\mathbf{x}_{i, 1:2(j-1)}), \quad
\mbox{and}\quad \phi_{ij0}=1-\phi_{ij1}-\phi_{ij2}
\label{eq-3}
\end{eqnarray}
which are conditional probabilities $\mbox{Pr}(\mbox{SNP}_{ji}=k|\mathbf{x}_{1:2(j-1)})$ with $k=1,2,0$, 
respectively. Here $\mathbf{x}_{i,1:2(j-1)}$ is the vector of the first $2(j-1)$ elements of $\mathbf{x}_i$ 
(Note $\mathbf{x}_{i, 1:0}=\emptyset$). Writing $\pmb{\phi}_{jk}=(\phi_{1jk},\cdots,\phi_{njk})^\top$
with $k=0,1,2$, we can naturally propose a trinomial logistic regression model for these conditional probabilities
\begin{eqnarray}
\log\frac{\phi_{ij1}}{\phi_{ij0}}=(\mathbf{z}_i^\top,\mathbf{x}_{i,1:2(j-1)}^\top)\pmb{\alpha}_{j1} 
\quad \Longleftrightarrow\quad \log\frac{\pmb{\phi}_{j1}}{\pmb{\phi}_{j0}}=
(Z, X_{1:2(j-1)})\pmb{\alpha}_{j1}, \nonumber \\
\log\frac{\phi_{ij2}}{\phi_{ij0}}=(\mathbf{z}_i^\top,\mathbf{x}_{i,1:2(j-1)}^\top)\pmb{\alpha}_{j2} 
\quad \Longleftrightarrow\quad \log\frac{\pmb{\phi}_{j2}}{\pmb{\phi}_{j0}}=
(Z, X_{1:2(j-1)})\pmb{\alpha}_{j2}, \quad j=1,\cdots, p,
\label{eq-4}
\end{eqnarray}
where $X_{1:2(j-1)}$ is the first $2(j-1)$ columns of $X$, and $\pmb{\alpha}_{jk}$ with $k=1,2$ is a 
$[2(q+j)-1]\times 1$ 
unknown parameter vector giving the associations of $\mbox{SNP}^1, \cdots, \mbox{SNP}^q$ and
$\mbox{SNP}_1, \cdots, \mbox{SNP}_{j-1}$ with $\mbox{SNP}_j$ and is to be estimated from the data.
Denote $\pmb{\alpha}_k=(\pmb{\alpha}_{1k}^\top,\cdots, \pmb{\alpha}_{pk}^\top)^\top$ with $k=1,2$.
By (\ref{eq-3}) and (\ref{eq-4}) the joint probability function of $\mathbf{x}_i$ equals
\begin{equation}
p_2(\mathbf{x}_i|\mathbf{z}_i, \pmb{\alpha}_1, \pmb{\alpha}_2)\!=\!\prod_{j=1}^p
p_2(x_{i,2j\!-\!1}, x_{i,2j}|\mathbf{z}_i, \mathbf{x}_{i, 1:2(j\!-\!1)}, \pmb{\alpha}_{j1}, \pmb{\alpha}_{j2})
\!=\!\!\prod_{j=1}^p\phi_{ij0}^{1-x_{i,2j\!-\!1}-x_{i,2j}}\phi_{ij1}^{x_{i,2j\!-\!1}}\phi_{ij2}^{x_{i,2j}}.
\label{eq-5}
\end{equation}
And by (\ref{eq-5}) the joint probability function of $X$ given $Z$ is
\begin{equation}
p_2(X|Z,\pmb{\alpha}_1, \pmb{\alpha}_2)=
\prod_{i=1}^n p_2(\mathbf{x}_i|\mathbf{z}_i, \pmb{\alpha}_1, \pmb{\alpha}_2)=\prod_{i=1}^n
\prod_{j=1}^p\phi_{ij0}^{1-x_{i,2j\!-\!1}-x_{i,2j}}\phi_{ij1}^{x_{i,2j\!-\!1}}\phi_{ij2}^{x_{i,2j}}.
\label{eq-6}
\end{equation}
Since the missing values in the data may be non-ignorable, a joint probability distribution is needed to model
the missingness mechanism. Namely, the joint conditional probability distribution of each missingness indicator
vector $\mathbf{r}_i$ given $(y_i, \mathbf{x}_i, \mathbf{z}_i)$ is needed, where each $r_{ij}$ follows a
Bernoulli distribution with probability $\psi_{ij}=\mbox{Pr}(r_{ij}=1|y_i,\mathbf{x}_i, \mathbf{z}_i)$ with
$j=1,\cdots,p$ and $i=1,\cdots,n$. By \citeA{ibrahim1999missing} this joint conditional probability function can be
expressed as the product of a sequence of univariate conditional probability functions, as given below
\begin{equation}
p_3(\mathbf{r}_i|\mathbf{z}_i,\mathbf{x}_i,y_i,\pmb{\gamma})=\prod_{j=1}^p
p_3(r_{ij}|\mathbf{r}_{i, 1:(j-1)}, \mathbf{z}_i,\mathbf{x}_i,y_i,\pmb{\gamma}_j)
\label{eq-7}
\end{equation}
where $\mathbf{r}_{i, 1:(j-1)}$ is the first $j-1$ elements of $\mathbf{r}_i$,
and $\pmb{\gamma}=(\pmb{\gamma}_1^\top, \cdots, \pmb{\gamma}_p^\top)^\top$ is an unknown
parameter vector specifying the parametric structure of $\{\pmb{\psi}_j=(\psi_{1j},\cdots, \psi_{nj})^\top,
j=1,\cdots,p\}$. It can be seen that $\dim(\pmb{\gamma}_j)=1+2q+2p+j$ and
$\dim (\pmb{\gamma})=p(1+2q+2p+\frac{p+1}{2})$. We propose to model the missingness probabilities
$\psi_{ij}$ by a set of logistic regression equations as given below
\begin{equation}
\log\frac{\psi_{ij}}{1\!-\!\psi_{ij}}\!=\!(\mathbf{z}_i^\top, \mathbf{x}_i^\top, y_i, \mathbf{r}_{i,1:(j\!-\!1)}^\top)
\pmb{\gamma}_j;\; i\!=\!1,\cdots\!, n \; \Longleftrightarrow\; \log\frac{\pmb{\psi}_j}{\mathbf{1}\!-\!\pmb{\psi}_j}
\!=\!(Z, X, \mathbf{y},R_{1:(j\!-\!1)})\pmb{\gamma}_j
\label{eq-8}
\end{equation}
for $j=1,\cdots, p$, where $\pmb{\gamma}_j$ gives the effect of $(Z, X, \mathbf{y},R_{1:(j-1)})$ on the
missingness mechanism of $\mbox{SNP}_j$ and $R_{1:(j-1)}$ is the first $j-1$ columns of $R$.
By (\ref{eq-7}) and (\ref{eq-8}) the joint conditional probability of $R$ given $(Z, X, \mathbf{y})$ is
\begin{eqnarray}
p_3(R|Z, X, \mathbf{y},\pmb{\gamma})\!&\!\!=\!\!&\! \prod_{i=1}^n
p_3(\mathbf{r}_i|\mathbf{z}_i,\mathbf{x}_i,y_i,\pmb{\gamma}) \nonumber \\
\!&\!\!=\!\!&\! \prod_{i=1}^n \prod_{j=1}^p 
p_3(r_{ij}|\mathbf{r}_{i, 1:(j-1)}, \mathbf{z}_i,\mathbf{x}_i,y_i,\pmb{\gamma}_j) \!=\!
\prod_{i=1}^n \prod_{j=1}^p \psi_{ij}^{r_{ij}}(1-\psi_{ij})^{1-r_{ij}}.
\label{eq-9}
\end{eqnarray}

Equations (\ref{eq-1}), (\ref{eq-4}) and (\ref{eq-8}) together with(\ref{eq-2}), (\ref{eq-6}) and (\ref{eq-9}) 
provide a system of logistic regression models for 
analyzing the complete data $(\mathbf{y}, Z, X, R)$, from which the phenotype-SNPs association can be inferred.
To perform such inference we first need to estimate the unknown parameter vector 
$\pmb{\theta}=(\pmb{\beta}^\top, \pmb{\alpha}_1^\top, \pmb{\alpha}_2^\top, \pmb{\gamma}^\top)^\top$
by method of maximum likelihood. From (\ref{eq-2}), (\ref{eq-6}) and (\ref{eq-9}) it is easy to see the join
log-likelihood function of $\pmb{\theta}$ given the complete data $(\mathbf{y}, Z, X, R)$ is
\begin{eqnarray}
\ell_c(\pmb{\theta})&=&\sum_{i=1}^n\ell_{ci}(\pmb{\theta})=\log p_1(\mathbf{y}|Z, X, \pmb{\beta})
+\log p_2(X|Z,\pmb{\alpha}_1, \pmb{\alpha}_2) + \log p_3(R|Z, X, \mathbf{y},\pmb{\gamma}) \nonumber \\
&=&\sum_{i=1}^n\left\{y_i\log\pi_i+(1-y_i)\log(1-\pi_i)+\sum_{j=1}^p\left[(1-x_{i,2j-1}-x_{i,2j})
\log\phi_{ij0}\right.\right. \nonumber \\
& & \hspace*{-20pt}\left.\phantom{\sum_{j=1}^p}\left.+x_{i,2j-1}\log\phi_{ij1}+x_{i,2j}\log\phi_{ij2}
+r_{ij}\log\psi_{ij}+(1-r_{ij})\log(1-\psi_{ij})\right]\right\}
\label{eq-10}
\end{eqnarray}
where $\ell_{ci}(\pmb{\theta})$ is the complete data log-likelihood for individual $i$. However, the maximum
likelihood estimator (MLE) of $\pmb{\theta}$ cannot be obtained by maximizing $\ell_c(\pmb{\theta})$,
because $\ell_c(\pmb{\theta})$ contains missing values. On the other hand, it is not feasible to compute
the MLE of $\pmb{\theta}$ through maximizing the observed data log-likelihood 
$\ell_{\rm obs}(\pmb{\theta})=\log\sum_{X_{\rm mis}} e^{\ell_c(\pmb{\theta})}$, with the summation $\sum_{X_{\rm mis}}$ being taken
over all possible values of the missing part $X_{\rm mis}$ of $X$, due to the involved mathematical
intractability and the fact that the $\mbox{SNP}_j$'s in $\mathbf{x}$ may be high-dimensional and in 
the state of linkage disequilibrium (LD).

To overcome this difficulty we propose to use EM algorithm \cite{Dempster1977EM} and the ridge regression
idea \cite{hoerl1981ridge} to compute $\hat{\pmb{\theta}}$, a penalized maximum likelihood estimator (PMLE)
of $\pmb{\theta}$, i.e.
\begin{equation}
\hat{\pmb{\theta}}= \underset{\pmb{\theta}}{\arg\!\max} \left\{\ell_{\rm obs}(\pmb{\theta})
-\frac{1}{2}\left[\lambda_1||\pmb{\beta}||_2^2 +\lambda_2(||\pmb{\alpha}_1||_2^2
+||\pmb{\alpha}_2||_2^2)+\lambda_3||\pmb{\gamma}||_2^2\right]\right\}
\label{eq-11}
\end{equation}
where $||\cdot ||_2$ is the $L_2$-norm, and $(\lambda_1, \lambda_2, \lambda_3)$ are tuning parameters that 
may be optimized by 
cross-validation or BIC to be detailed later in the paper. To reduce the computing complexity we may
set $\lambda_1=\lambda_2=\lambda_3=\lambda$.
For given value of $(\lambda_1, \lambda_2, \lambda_3)$, PMLE $\hat{\pmb{\theta}}$ is computed
by the following Ridge-EM algorithm. \vspace*{-8pt}

\paragraph{Algorithm 1}  Computing PMLE by
ridge-expectation-maximization (\textbf{Ridge-EM}):
\begin{itemize}
\item[$1^\circ$]
Set a plausible initial estimate $\pmb{\theta}^{(0)}$ for $\pmb{\theta}$.
\item[$2^\circ$]
Repeat for $k=0,1,\cdots$ until convergence of computation:
\begin{itemize}
\item[$2.1^\circ$] \textbf{E-step}
Compute 
\begin{eqnarray}
Q(\pmb{\theta}|\pmb{\theta}^{(k)}) &=& 
E\left[\ell_c(\pmb{\theta})|X_{\rm obs}, Z, \mathbf{y}, R, \pmb{\theta}^{(k)}\right] \nonumber \\
& & -\frac{1}{2}\left[\lambda_1||\pmb{\beta}||_2^2 +\lambda_2(||\pmb{\alpha}_1||_2^2
+||\pmb{\alpha}_2||_2^2)+\lambda_3||\pmb{\gamma}||_2^2\right]
\label{eq-12}
\end{eqnarray}
where the first term is the expectation of $\ell_c(\pmb{\theta})$ with respect to the conditional distribution of
$X_{\rm mis}$ given the observed data. Here $X_{\rm obs}$ and $X_{\rm mis}$ are the observed and the
missing part of $X$, respectively. Note that the parameter value is $\pmb{\theta}^{(k)}$ in the conditional
distribution of $X_{\rm mis}$.
\item[$2.2^\circ$] \textbf{M-step}
Find $\pmb{\theta}^{(k+1)}=\underset{\pmb{\theta}}{\arg\!\max}\, Q(\pmb{\theta}|\pmb{\theta}^{(k)})$
by solving $\displaystyle \frac{\partial Q(\pmb{\theta}|\pmb{\theta}^{(k)})}{\partial \pmb{\theta}}=\mathbf{0}$.
\end{itemize}
\item[$3^\circ$]
The PMLE $\displaystyle \hat{\pmb{\theta}}=\lim_{k\rightarrow\infty}\pmb{\theta}^{(k)}$. By the method
of \citeA{Louis1982}, we estimate $\mbox{var}(\hat{\pmb{\theta}})$ by
\begin{equation}
\widehat{\mbox{var}}(\hat{\pmb{\theta}})=\left[-\frac{\partial^2 Q(\pmb{\theta}|\hat{\pmb{\theta}})}{
\partial \pmb{\theta}\partial \pmb{\theta}^\top}-\mbox{var}\left(\left.\frac{\partial \ell_c(\pmb{\theta})}{\partial
\pmb{\theta}}\right| X_{\rm obs}, Z, \mathbf{y}, R, \hat{\pmb{\theta}}\right) \right]_{\pmb{\theta}=
\hat{\pmb{\theta}}}^{-1}
\label{eq-13}
\end{equation}
\end{itemize}

It is not difficult to show that the estimate obtained from the Ridge-EM algorithm is indeed the PMLE
under general regularity conditions.
However, a full implementation of the Ridge-EM algorithm until convergence would be computationally very
intensive and actually not necessary, when many redundant predictor variables are still not removed from
the logistic regression system (\ref{eq-1},~\ref{eq-4},~\ref{eq-8}) in use. We suggest to fully implement
the Ridge-EM algorithm only at the final stage when all insignificant predictors are eliminated from the system.
Before this final stage, we propose a missing value multiple imputation method to simplify the Ridge-EM
algorithm.

For $Q(\pmb{\theta}|\pmb{\theta}^{(k)})$ in the E-step it can be shown that 
\begin{eqnarray}
Q(\pmb{\theta}|\pmb{\theta}^{(k)})\!\!&\!\!=\!\!&\!\! \sum_{i=1}^n\!E\!\left[\ell_{ci}(\pmb{\theta})|
\mathbf{x}_{i, {\rm o}},\mathbf{z}_i, y_i, \mathbf{r}_i, \pmb{\theta}^{(\!k\!)}\right]
\!-\!\frac{1}{2}\!\left[\!\lambda_1||\pmb{\beta}||_2^2 \!+\!\!\lambda_2(||\pmb{\alpha}_1||_2^2
\!+\!\!||\pmb{\alpha}_2||_2^2)\!+\!\!\lambda_3||\pmb{\gamma}||_2^2\right] \nonumber \\
\!\!&\!\!=\!\!&\!\! \sum_{i=1}^n\sum_{\mathbf{x}_{i, {\rm m}}}\omega_{i, {\rm m}}^{(k)}
\left[\log p_1(y_i|\mathbf{z}_i,\mathbf{x}_{i,{\rm o}}, \mathbf{x}_{i,{\rm m}}, \pmb{\beta})
\right. \nonumber \\
& & \quad\quad \quad
\left. +\log p_2(\mathbf{x}_{i,{\rm o}}, \mathbf{x}_{i,{\rm m}}|\mathbf{z}_i, \pmb{\alpha}_1, 
\pmb{\alpha}_2) +\log p_3(\mathbf{r}_i|\mathbf{z}_i,\mathbf{x}_{i,{\rm o}}, \mathbf{x}_{i,{\rm m}},
y_i,\pmb{\gamma})\right] \nonumber \\
& & \quad\quad\quad -\frac{1}{2}\left[\lambda_1||\pmb{\beta}||_2^2 +\lambda_2(||\pmb{\alpha}_1||_2^2
+||\pmb{\alpha}_2||_2^2)+\lambda_3||\pmb{\gamma}||_2^2\right]
\label{eq-14}
\end{eqnarray}
where $\mathbf{x}_{i,{\rm o}}$ and $\mathbf{x}_{i,{\rm m}}$ are the observed and missing parts of
$\mathbf{x}_i$, respectively. Thus, $(\mathbf{x}_{i,{\rm o}}, \mathbf{x}_{i,{\rm m}})=\mathbf{x}_i$
subject to a permutation of $\mathbf{x}_i$. Here $\sum_{\mathbf{x}_{i, {\rm m}}}$ means the summation
over all $3^{\nu_i}$ possible values that $\mathbf{x}_{i, {\rm m}}$ can take, where $\nu_i$ denotes
the number of SNP variables with their values for individual $i$ being missing (clearly $0\leq\nu_i\leq p$).
Each term $\omega_{i, {\rm m}}^{(k)}$ in (\ref{eq-14}) is the conditional probability
$\mbox{Pr}(\mathbf{x}_{i,{\rm m}}|\mathbf{z}_i,\mathbf{x}_{i,{\rm o}}, y_i, \mathbf{r}_i,
\pmb{\theta}^{(k)})$, i.e.
\begin{equation}
\omega_{i, {\rm m}}^{(k)}=\frac{p_1(y_i|\mathbf{z}_i,\mathbf{x}_{i,{\rm o}}, \mathbf{x}_{i,{\rm m}},
\pmb{\beta}^{(k)})p_2(\mathbf{x}_{i,{\rm o}}, \mathbf{x}_{i,{\rm m}}|\mathbf{z}_i, \pmb{\alpha}_1^{(k)}, 
\pmb{\alpha}_2^{(k)})p_3(\mathbf{r}_i|\mathbf{z}_i,\mathbf{x}_{i,{\rm o}}, \mathbf{x}_{i,{\rm m}},
y_i,\pmb{\gamma}^{(k)})}{\sum_{\mathbf{x}_{i, {\rm m}}^\prime}
p_1(y_i|\mathbf{z}_i,\mathbf{x}_{i,{\rm o}}, \mathbf{x}_{i,{\rm m}}^\prime,\pmb{\beta}^{(k)})
p_2(\mathbf{x}_{i,{\rm o}}, \mathbf{x}_{i,{\rm m}}^\prime|\mathbf{z}_i, \pmb{\alpha}_1^{(k)}, 
\pmb{\alpha}_2^{(k)})p_3(\mathbf{r}_i|\mathbf{z}_i,\mathbf{x}_{i,{\rm o}}, \mathbf{x}_{i,{\rm m}}^\prime,
y_i,\pmb{\gamma}^{(k)})}
\label{eq-15}
\end{equation}
If we regard the $3^{\nu_i}$ possible values that each $\mathbf{x}_{i, {\rm m}}$ can take as the
$3^{\nu_i}$ imputations for $\mathbf{x}_{i,{\rm m}}$, which together with the observed data
constitute the complete data (for individual $i$), then $Q(\pmb{\theta}|\pmb{\theta}^{(k)})$ in (\ref{eq-14})
can be seen as a penalized weighted log-likelihood function of $\pmb{\theta}$ for the complete data.
Therefore, $\pmb{\theta}^{(k+1)}$ computed from the M-step is just the maximum penalized weighted
log-likelihood estimate (MPWLE) that can be computed from fitting the logistic regression system 
(\ref{eq-1},~\ref{eq-4},~\ref{eq-8}) with the complete data. 

Clearly, $\pmb{\theta}^{(k+1)}$ depends on $\lambda_1, \lambda_2$ and $\lambda_3$, tuning of which
for optimal model estimation may be computationally intensive. Therefore, we set $\lambda_1=\lambda_2
=\lambda_3=\lambda$ and tune only $\lambda$ in Ridge-EM. This should have little impact on optimal
model estimation because the predictors involved in (\ref{eq-1},~\ref{eq-4},~\ref{eq-8}) are of the similar
scale and magnitude. Tuning for best $\lambda$ value could be carried out by cross-validation, which in fact
is difficult to implement in practice. This is because the missingness pattern of the predictors in a
training sample may be different from that in a validation sample in cross-validation.  
A consequence of such variation in the missingness pattern is that fitting the model system 
(\ref{eq-1},~\ref{eq-4},~\ref{eq-8}) is likely to encounter singularity, in that some missingness indicator
variable $r_j$ has no variation across the training sample although $r_j$ varies across all samples.
Hence, modification is needed to avoid such singularity if one still chooses to use cross-validation.
For example, generalized cross-validation (GCV) of \citeA{GCV78} may be used. In this paper we propose
to use a BIC \cite{BIC78} based approach for tuning $\lambda$. Knowing that 
$\pmb{\theta}^{(k+1)}=\arg\!\max_{\pmb{\theta}}Q(\pmb{\theta}|\pmb{\theta}^{(k)})$
with $Q(\pmb{\theta}|\pmb{\theta}^{(k)})$ being computed by (\ref{eq-14}), a technique of the partition distribution of the parameter space \cite{QianWuXu2019} can be used to derive an empirical BIC as
\begin{equation}
\mbox{EBIC}(\lambda)=-2Q(\pmb{\theta}^{(k+1)}|\pmb{\theta}^{(k)}) - \mbox{trace}\left(\!\!
\widehat{\mbox{var}}(\pmb{\theta}^{(k+1)})\!\cdot \!\left[\frac{\partial^2 Q(\pmb{\theta}|\pmb{\theta}^{(k)})}{
\partial \pmb{\theta}\partial \pmb{\theta}^\top}\right]_{\pmb{\theta}=\pmb{\theta}^{(k+1)}}\right)\log((2p+1)n)^\xi
\label{eq-16}
\end{equation}
where $Q(\pmb{\theta}^{(k+1)}|\pmb{\theta}^{(k)}) $ is given by (\ref{eq-14}) with $\lambda_1=\lambda_2
=\lambda_3=\lambda$; $\widehat{\mbox{var}}(\pmb{\theta}^{(k+1)})$ is from (\ref{eq-13}); and $\xi$, with the default value 2.0,
is an adjustment parameter tuning the effective volume 
of the sample space. Note that $\lambda$ is implicitly
involved in each term in (\ref{eq-16}).
Given a pre-specified set of $\lambda$ values, the best $\lambda$ value used in Ridge-EM is the one
minimizing EBIC($\lambda$). We have used the \texttt{glmnet} package in \texttt{R} to implement
the Ridge-EM algorithm together with tuning $\lambda$ by EBIC.

Another complication in computing the MPWLE $\pmb{\theta}^{(k+1)}$ is the value of 
$\pmb{\gamma}^{(k+1)}$ could be overly biased, especially when some column(s) of $R$, the
observed data matrix of the missingness indicator vector $\mathbf{r}$, are imbalanced in that there are
either too few 1's or two few 0's. If this is the case, we choose to use the bias reduction method of
\citeA{firth1993bias} to compute the MPWLE $\pmb{\theta}^{(k+1)}$ in the M-step of Ridge-EM.
This bias reduction method has been implemented in \texttt{R} package \texttt{brglm} 
\cite{kosmidis2020package}.

Since a key objective in GWAS is about SNP selection from logistic regression model (\ref{eq-1}),
and it may be computationally very expensive to perform parameter estimation in SNP selection by executing
the Ridge-EM algorithm in iteration until convergence, we decide to apply the Ridge-EM algorithm
for small $\kappa$ iterations (e.g. $\kappa\leq 5$) to only generate all possible values that
each $\mathbf{x}_{i,{\rm m}}$ can take and the corresponding $3^{\nu_i}$ conditional
probabilities $\omega_{i, {\rm m}}^{(k)}$, for $i=1,\cdots, n$. We name this work the 
\textit{missing value multiple imputations}. We then apply a random forest (RF) method to perform
SNPs variables ($\mathbf{z}$ and $\mathbf{x}$) selection by the \texttt{VIMP} measure for each
equation in system (\ref{eq-1},~\ref{eq-4},~\ref{eq-8}). Note that variable selection from (\ref{eq-8})
also involves $\mathbf{r}_{1:(p-1)}$ and $y$. The details
will be given in section~\ref{sec-2.2}.

Next, we remove those unimportant SNPs and other predictors determined by RF from each equation in system
(\ref{eq-1},~\ref{eq-4},~\ref{eq-8}), and perform missing value multiple imputations again
by applying the Ridge-EM algorithm with $\kappa$ iterations on the updated (\ref{eq-1},~\ref{eq-4},~\ref{eq-8}).
This procedure can be repeatedly performed until
variable selection by RF stabilizes, and accordingly the selected model system
(\ref{eq-1},~\ref{eq-4},~\ref{eq-8}) stabilizes. Once this stabilization is achieved, we apply
the Ridge-EM algorithm (or the EM algorithm) to convergence to compute the PMLE
(or MLE) of $\pmb{\theta}_{\rm final}$, the vector parameter in the stabilized model
system (\ref{eq-1},~\ref{eq-4},~\ref{eq-8}).

Finally, note that generating all possible values of $\mathbf{x}_{i,{\rm m}}$ with conditional probabilities $\omega_{i, {\rm m}}^{(k)}$
is computationally infeasible when $3^{\nu_i}$ is very large. In this situation, we can use Gibbs sampler
to randomly generate a sample of $\mathbf{x}_{i,{\rm m}}$ values of size $l_i$. Then 
$Q(\pmb{\theta}|\pmb{\theta}^{(k)})$ given in (\ref{eq-14}) can be approximated (to any pre-specified
precision by tuning $l_i$) by replacing $\omega_{i, {\rm m}}^{(k)}$ with $l_i^{-1}$ and
replacing $\sum_{\mathbf{x}_{i,{\rm m}}} \cdots$ in (\ref{eq-14}) with the sample average.
It is easy to see that each of the $\nu_i$ univariate conditional distributions used in the Gibbs sampler here is a
trinomial distribution. In the situation where $3^{\nu_i}$ is large but not very large so that enumerating
$(\mathbf{x}_{i,{\rm m}}, \omega_{i, {\rm m}}^{(k)})$ is computationally feasible, we can still approximate
$Q(\pmb{\theta}|\pmb{\theta}^{(k)})$ in (\ref{eq-14}) with sufficient precision by replacing those
small $\omega_{i, {\rm m}}^{(k)}$ probabilities with 0, and normalizing the remaining 
$\omega_{i, {\rm m}}^{(k)}$ values by reweighing. This approximation may substantially reduce the
computing intensity.

\subsection{Random forest for variable selection and elimination} \label{sec-2.2}
Since it is computationally infeasible to perform variable selection based on parametric equations
(\ref{eq-1},~\ref{eq-4},~\ref{eq-8}) when $p+q$ is large, as discussed in the section~\ref{sec-2.1}, we propose
to apply RF to select the most important or eliminate the least important predictor variables regarding their effects
on the response variables in equations (\ref{eq-1},~\ref{eq-4},~\ref{eq-8}). The computing efficiency of RF
for high-dimensional genetic variable selection with highly correlative variables is proven, cf. \citeA{diaz2006gene}
and the review of \citeA{chen2012random}, whereas the selection and elimination itself is performed based on
certain \texttt{VIMP} measure.

There are two common \texttt{VIMP} measures for each predictor variable in RF. The first is derived based
on the mean decrease of the model prediction accuracy (MDA) when the predictor is excluded from prediction
for OOB samples. The MDA measure of \texttt{VIMP} is also named permutation importance (or Breiman-Cutler
importance). It can be seen that a predictor is more important when its MDA value is larger. The second
importance measure is based on the mean decrease of Gini impurity (MDG) over those nodes in the RF
where a predictor under consideration is chosen to make the split.

While the MDG measure for each predictor is automatically computed during the process of RF fitting, the MDA
importance measure can be computed with little extra effort. Specifically, to compute the MDA value of a predictor
$x$ on the categorical response variable in an RF, one proceeds with the following procedure.

\paragraph{Algorithm 2}  Computing MDA for each predictor in RF:
\begin{itemize}
\item[$1^\circ$]
Suppose there are \texttt{ntr} trees $\{\mathcal{T}_l, l=1,\cdots, \texttt{ntr}\}$ and \texttt{mtr} predictors
$\{x^{(j)}, j=1,\cdots, \texttt{mtr}\}$ in the RF.
Use each tree $\mathcal{T}_l$ to predict the values of the response (denoted as $\mathbf{y}_l$) of the OOB
sample individuals associated with $\mathcal{T}_l$. Denote as $\mbox{OOB}_l$ the collection of all predictors' 
values for the OOB sample individuals in $\mathcal{T}_l$, and as $\hat{\mathbf{y}}_l$ the response's predictions at
$\mbox{OOB}_l$. Compute the total prediction error $L(\hat{\mathbf{y}}_l, \mathbf{y}_l)$,
where $L(\cdot, \cdot)$ is a loss function, e.g. mis-classification error.
\item[$2^\circ$]
Repeat for each predictor $x^{(j)}$, $j=1,\cdots, \texttt{mtr}$; and for each $j$ repeat for each $\mathcal{T}_l$,
$l=1,\cdots, \texttt{ntr}$:
\begin{itemize}
\item[$2.1^\circ$]
Denote $|\mbox{OOB}_l|$ as the number of individuals in the OOB sample corresponding to $\mathcal{T}_l$.
Also denote the set of all values of $x^{(j)}$ in $\mbox{OOB}_l$ as $\mathbf{x}^{(j)}_l$.
\item[$2.2^\circ$]
Update $\mbox{OOB}_l$ to $\widetilde{\mbox{OOB}}_l$ by replacing $\mathbf{x}^{(j)}_l$ with
$\tilde{\mathbf{x}}^{(j)}_l$, a permutation of $\mathbf{x}^{(j)}_l$.
\item[$2.3^\circ$]
Use $\mathcal{T}_l$ to predict the response at $\widetilde{\mbox{OOB}}_l$ . 
Denote the predictions as $\hat{\mathbf{y}}_l^{(j)}$.
\item[$2.4^\circ$]
Compute $L(\hat{\mathbf{y}}_l^{(j)}, \mathbf{y}_l)$, the total prediction error at $\widetilde{\mbox{OOB}}_l$.
\end{itemize}
\item[$3^\circ$]
The MDA importance value of $x^{(j)}$ is computed as
\begin{equation}
\mbox{MDA}(x^{(j)})=\frac{1}{\texttt{ntr}}\sum_{l=1}^{\texttt{ntr}}\frac{L(\hat{\mathbf{y}}_l^{(j)}, 
\mathbf{y}_l)-L(\hat{\mathbf{y}}_l, \mathbf{y}_l)}{|\mbox{OOB}_l|}, \quad j=1,\cdots, \texttt{mtr}.
\label{eq-17}
\end{equation}
\end{itemize}
It is not difficult to see that each $\mbox{MDA}(x^{(j)})$ mostly has a positive value and tends to be large
if $x^{(j)}$ has important effect on the response. On the other hand, $\mbox{MDA}(x^{(j)})$ could take
a negative value if $x^{(j)}$ is not important for growing the RF, e.g. when $x^{(j)}$ is a noise variable.

After multiple imputations of missing values are generated using the Ridge-EM algorithm, a set of complete
data will be formed by combining the imputations and the observed data, properly weighted according to the
conditional probabilities $\omega_{i, {\rm m}}^{(k)}$ given in (\ref{eq-15}). Then for each equation in
(\ref{eq-1},~\ref{eq-4},~\ref{eq-8}) an RF can be grown using the complete data together with the weights 
$\omega_{i, {\rm m}}^{(k)}$ in (\ref{eq-15}), and a \texttt{VIMP} measure such as MDA or MDG can be
computed for each predictor variable (mostly a SNP variable, may also be the phenotype $y$ or a missingness
indicator $r_j$) appeared in (\ref{eq-1},~\ref{eq-4},~\ref{eq-8}).

Based on the computed \texttt{VIMP} values, for each RF linking with an equation in the system 
(\ref{eq-1},~\ref{eq-4},~\ref{eq-8}) we can identify $s$
predictors having the $s$ highest \texttt{VIMP} values in that RF, with $s$ being a pre-specified
integer. Alternatively, we can identify those predictors in each RF that have their \texttt{VIMP} values
greater than a pre-specified threshold value. Then we include only those predictors selected by each RF into the
corresponding equation from (\ref{eq-1},~\ref{eq-4},~\ref{eq-8}), with all 
other predictors being eliminated from that equation, so that all equations in (\ref{eq-1},~\ref{eq-4},~\ref{eq-8}) 
are updated. Next we repeat the multiple imputation step by applying the Ridge-EM algorithm to the updated system
(\ref{eq-1},~\ref{eq-4},~\ref{eq-8}), so that new complete data are formed. We continue this 
\textit{Ridge-EM $\rightarrow$ multiple imputations $\rightarrow$ complete data $\rightarrow$ RF
$\rightarrow$ variable selection $\rightarrow$ Ridge-EM} cycle until computational stability is achieved.
In practice, this cycle is repeated for only a pre-specified number $\tau$ times, e.g. $\tau=10$,
regardless of whether computational convergence has been achieved.

To mitigate the possible adverse effect of non-convergence, the final variable selection decision will
be made based on the frequencies of the predictors selected over the $\tau$ cycles of the RF fitting. Namely,
for each equation in (\ref{eq-1},~\ref{eq-4},~\ref{eq-8}), only those predictions, having the pre-specified
$s^*$ highest frequencies in the $\tau$ iterations of the corresponding RF fitting, are included in the equations
at the end. Once this final variable selection is done to obtain the final form of the system 
(\ref{eq-1},~\ref{eq-4},~\ref{eq-8}), we perform the Ridge-EM on (\ref{eq-1},~\ref{eq-4},~\ref{eq-8})
in iteration until convergence, or fit (\ref{eq-1},~\ref{eq-4},~\ref{eq-8}) using the most recently formed
complete data, to find the PMLE or MPWLE $\hat{\pmb{\theta}}_{\rm final}$ and 
$\mbox{var}(\hat{\pmb{\theta}}_{\rm final})$ etc.. 

The following summarizes the procedure we developed here for variable selection and parameter estimation
for studying the phenotype-SNP associations involving non-ignorable missing values.

\paragraph{Algorithm 3}  Multiple imputations by Ridge-EM and variable selection from RF
(\textbf{MiREM-VSeRF}):
\begin{itemize}
\item[$1^\circ$]
Set a plausible initial estimate $\pmb{\theta}^{(0)}$ for $\pmb{\theta}$ in logistic system 
(\ref{eq-1},~\ref{eq-4},~\ref{eq-8}).
\item[$2^\circ$]
Compute an estimate update of $\pmb{\theta}$ using the Ridge-EM algorithm, where multiple imputations
for missing values are generated at the same time. Combine the generated multiple imputations, the
associated conditional probabilities (\ref{eq-15}), and the observed data to form a tentative set of complete data.
\item[$3^\circ$]
Use the tentative complete data to create a random forest for each equation in system 
(\ref{eq-1},~\ref{eq-4},~\ref{eq-8}). Then use a \texttt{VIMP} measure, e.g. MDA or MDG, to perform
variable selection from each RF. Accordingly, update each equation in (\ref{eq-1},~\ref{eq-4},~\ref{eq-8})
by including as predictors only the selected variables from the corresponding RF.
\item[$4^\circ$]
Repeat $2^\circ$ and $3^\circ$ a pre-specified number $\tau$ times, e.g. $\tau=10$.
\item[$5^\circ$]
Apply Ridge-EM to compute $\hat{\pmb{\theta}}_{\rm final}$, the PMLE of $\pmb{\theta}_{\rm final}$,
together with an estimated variance of $\hat{\pmb{\theta}}_{\rm final}$, based on the latest update of
system (\ref{eq-1},~\ref{eq-4},~\ref{eq-8}) from $4^\circ$.
\end{itemize}

\subsection{Inference about the missingness mechanism} \label{sec-2.3}
Advantages of using the logistic regression system (\ref{eq-1},~\ref{eq-4},~\ref{eq-8}) to study the 
phenotype-SNP associations include that equation (\ref{eq-8}) provides a venue to make inference about the
missingness mechanism of those SNP variables having missing values. This inference is pursued through
linear hypothesis testing.

For example, whether or not a sub-vector of $(\mathbf{z}^\top, \mathbf{x}^\top, y, \mathbf{r}_{1:(j-1)}^\top)$,
indexed by $\zeta$ --- a subset of $\{1,\cdots, \dim(\pmb{\gamma}_j)\}$, has significant effect on the missingness
indicator $r_j$ can be studied by testing the null hypothesis $H_0: \pmb{\gamma}_{j\zeta}=\mathbf{0}$
versus $H_1: \pmb{\gamma}_{j\zeta}\neq\mathbf{0}$ at a given significance level $\alpha$, where
$\pmb{\gamma}_{j\zeta}$ is a sub-vector of $\pmb{\gamma}_j$ indexed by $\zeta$. The Wald test statistic
$W=\hat{\pmb{\gamma}}_{j\zeta}^\top \left[\widehat{\mbox{var}}(\hat{\pmb{\gamma}}_{j\zeta})\right]^{-1}
\hat{\pmb{\gamma}}_{j\zeta}$, where $\hat{\pmb{\gamma}}_{j\zeta}$ is the PMLE computed from
Ridge-EM and $\widehat{\mbox{var}}(\hat{\pmb{\gamma}}_{j\zeta})$ is the estimated variance matrix of
$\hat{\pmb{\gamma}}_{j\zeta}$, asymptotically follows a $\chi^2_{\dim(\zeta)}$ distribution under $H_0$.
We can use $W$ and its asymptotic null distribution to compute the $p$-value of the test, and accordingly
draw a statistical conclusion about the missingness mechanism of each $\mbox{SNP}_j$ variable.

\section{Simulation study} \label{sec-3}
Two simulations are performed in this section to assess the performance of our proposed method consisting of
ridge regression, EM algorithm, multiple missing data imputation, random forecast and variable selection.
The first simulation considers the situation where the binary phenotype response is associated with a few
genotyped SNP predictors. The second one considers the situation where the binary phenotype response is 
associated with a moderate size of genotyped SNP predictors. In both simulations, we generate 100 SNP 
predictors and 1000 sample individuals, with the generated data containing non-ignorable missing values.
The data are then analyzed by the proposed method, with the results to be evaluated and the method justified.
The details are presented below.

\subsection{Genotyped SNP data generation} \label{sec-3.1}
\begin{figure}[!htbp]	
	\includegraphics[width=6.0in]{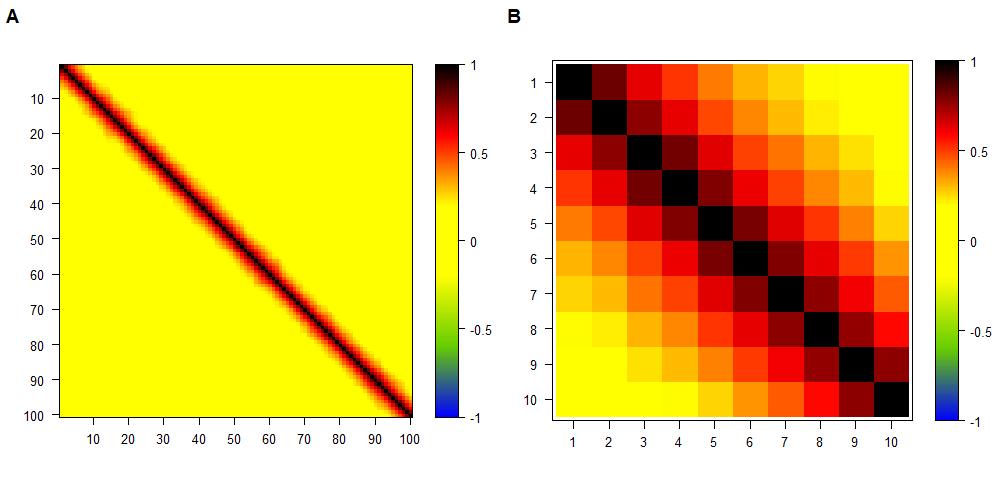}
	\centering
	\caption{\textit{Left panel} (\textbf{A}) is the correlation heat-map for the 100 simulated SNPs.
	\textit{Right panel} (\textbf{B}) displays the correlation heat-map for the first 10 simulated SNPs. }
	\label{Fig1}
\end{figure}

We generate a $1000\times 100$ matrix $G=(g_{i,j})_{i=1,\cdots,1000}^{j=1,\cdots, 100}$ to represent the
observations of 100 genotyped SNPs for 1000 individuals. That is, $g_{i,j}$ is the observation of $\mbox{SNP}_j$
(denoted as $g_j$) for individual $i$, and equals 0, 1, or 2. Accordingly, we use two dummy variables
$g_{i,j}^{(1)}$ and $g_{i,j}^{(2)}$ to encode $g_{i,j}$, i.e., $g_{i,j}^{(k)}=1$ if $g_{i,j}=k$ and
$g_{i,j}^{(k)}=0$ otherwise, $k=1$ or 2.
Also denote $g_j^{(1)}$ and $g_j^{(2)}$ (or $\mbox{SNP}_j^{(1)}$ and $\mbox{SNP}_j^{(2)}$) as the two dummy
variables for $\mbox{SNP}_j$ (or $g_j$) such that $g_j^{(k)}=1$ if $g_j=k$ and $g_j^{(k)}=0$ if $g_j\neq k$,
$k=1$ or 2. The generated values are obtained from running function \texttt{PopulationSNPSet()} in \texttt{R}
package \texttt{SNPSetSimulations} (\texttt{URL: //github.com/fhebert/SNPSetSimulations}), in which
the minor allele frequency (MAF) value for each SNP is randomly taken from \texttt{Uniform(0.3, 0.4)}
distribution, and the correlation coefficient matrix of the 100 SNPs is set to be
$$
R^*=
\begin{pmatrix}
1&0.8^{1}&0.8^2&0.8^3&\cdot&\cdot&\cdot&\cdot&\cdot&0.8^{100}\\
0.8^1&1&0.8^1&0.8^2&\cdot&\cdot&\cdot&\cdot&\cdot&0.8^{99}\\
0.8^2&0.8^1&1&0.8^1&\cdot&\cdot&\cdot&\cdot&\cdot&0.8^{98}\\
\cdot&\cdot&\cdot&\cdot&\cdot&\cdot&\cdot&\cdot&\cdot&\cdot\\
\cdot&\cdot&\cdot&\cdot&\cdot&\cdot&\cdot&\cdot&\cdot&\cdot\\
\cdot&\cdot&\cdot&\cdot&\cdot&\cdot&\cdot&\cdot&\cdot&\cdot\\
0.8^{99}&0.8^{98}&0.8^{97}&\cdot&\cdot&\cdot&\cdot&\cdot&1&0.8^{1}\\
0.8^{100}&0.8^{99}&0.8^{98}&\cdot&\cdot&\cdot&\cdot&\cdot&0.8^{1}&1
\\
\end{pmatrix}
$$
Figure~\ref{Fig1} left panel displays the heatmap of $R$ with the right panel being the part of that for the
first 10 SNPs.

\subsection{Simulation study 1} \label{sec-3.2}
\begin{figure}[!htbp]	
		\includegraphics[width=6in]{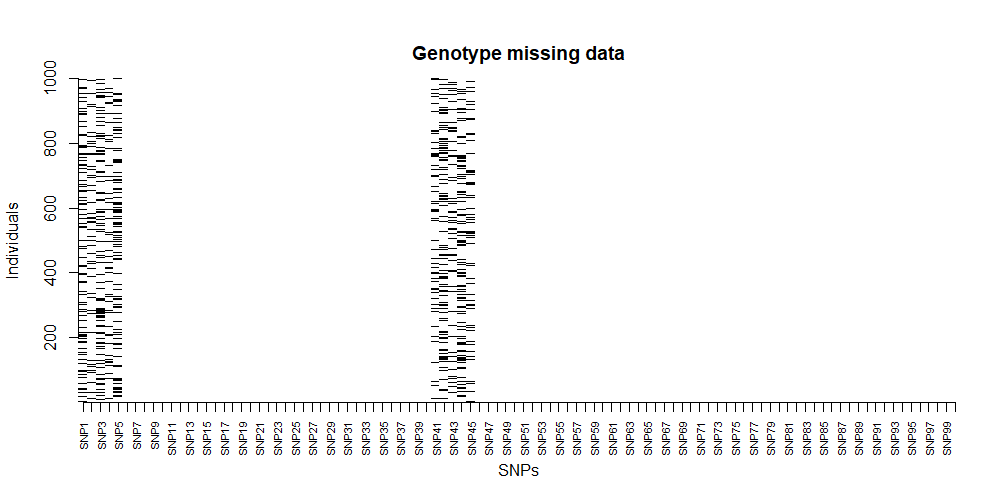}
		\centering
	\caption{Missingness patterns for SNPs in Simulation 1. Black bar at a (SNP, individual) location indicates the
	corresponding SNP value is missing for the referenced individual. Blank bar indicates it is observed.}
	\label{Fig1-sim1}
	\end{figure}

Consider the situation where the phenotype variable is associated with only a few SNPs, the sparse association
scenario. We generate 1000 values of the phenotype variable $y$ according to the following logistic model

\begin{equation}
\log\frac{\pi_i^*}{1-\pi_i^*}
=-2.2+1.6g_{i,41}^{(1)}+1.4g_{i,41}^{(2)}+1.8g_{i,42}^{(1)}-0.8g_{i,42}^{(2)}-1.7g_{i,50}^{(1)}-0.9g_{i,50}^{(2)}
	\label{eq-18}
\end{equation}
where $\pi_i^*=\mbox{Pr}(y_{i}=1)$, $i=1,\cdots, 1000$. We further remove some observations of $g_1,\cdots, g_5$,
$g_{41},\cdots, g_{45}$ to make them missing according to the following logistic model for missingness mechanism
\begin{equation}
\frac{\mbox{Pr}(r_{ij}\!=\!1)}{\mbox{Pr}(r_{ij}\!=\!0)}\equiv\log\frac{\psi_{ij}^*}{1-\psi_{ij}^*}
=\xi_{j,0}^*+\xi_{j,1}^*g_{i,j};\;j=1,\cdots,5,41,\cdots\!,45;\; i=1,\cdots\!, 1000
	\label{eq-19}
\end{equation}
with $g_{i,j}$ being the $i$th observation of $g_j$, and the values of $\xi_{j,0}^*$ and $\xi_{j,1}^*$ being given
in Table~\ref{tab-1}. Therefore, the missingness mechanism underpinning these missing values is non-ignorable
for the following statistical analysis. Proportions of missing values in the resultant observations of
$\mbox{SNP}_1, \cdots, \mbox{SNP}_5$ and $\mbox{SNP}_{41},\cdots,\mbox{SNP}_{45}$ are also presented in Table~\ref{tab-1} and summarized in Figure~\ref{Fig1-sim1}.
It can be found from the simulated data of $\mbox{SNP}_1,\cdots, \mbox{SNP}_5, \mbox{SNP}_{41}, \cdots, 
\mbox{SNP}_{45}$ that the total number of missing values in the data is 1958 and the numbers of 
individuals having 0,1,2,3,4,5 and 6 missing values are 114,294,283,179,93,33 and 4, respectively.

\renewcommand{\arraystretch}{1.1}
\begin{table}[!htbp]
		\centering
	\caption{Parameter values used in the missingness mechanism model and the resultant proportions of missing
	SNP values in Simulation 1.} \vspace{5pt}
	\begin{tabular}{|c|cccccccccc|} \hline
		$\mbox{SNP}_j$ & $g_1$ & $g_2$ & $g_3$ & $g_4$ & $g_5$ & $g_{41}$ & $g_{42}$ & $g_{43}$ & $g_{44}$ & $g_{45}$ \\
		\hline
		$\xi_{j,0}^*$ & $-2.0$ & $-2.0$ & $-2.0$ & $-2.0$ & $-2.0$ & $-2.0$ & $-2.0$ & $-2.0$ & $-2.0$ & $-2.0$  \\
		$\xi_{j,1}^*$ & $1.1$ & $0.4$ & $1.1$ & $0.4$ & $1.1$ & $0.4$ & $1.1$ & $0.4$ & $1.1$ & $0.4$  \\
		\%missing & .221 & .144 & .220 & .151 & .249 & .151 & .267 & .150 & .234 & .171 \\ \hline
	\end{tabular}
\label{tab-1}
\end{table}

With the generated data of $G$, $\mathbf{y}$, missing values and missingness mechanism, we analyze the 
phenotype-SNP association first by some traditional statistical methods ignoring the missing values.
This analysis consists of three parts:
\begin{itemize}
    \item[A1]
    Use the complete SNPs data (before creating the missing data) and the phenotype data
    to perform both random forecast variable selection and the logistic regression analysis based on
    all the 100 SNPs. The results (restricted to present only that of $\mbox{SNP}_{41}, \mbox{SNP}_{42}$ 
    and $\mbox{SNP}_{50}$ for brevity) are summarized in the top panel of Table~\ref{tab-2}, where each 
    $p$-value is for testing the significance of the corresponding $g_j^{(k)}$ in the logistic regression 
    analysis, and each MDA is for the corresponding $g_j$ in the random forecast analysis.
    \item[A2]
    Same as A1, except that the analysis is based on those 114 individuals not having any missing
    values. The other 886 individuals who each has at least one missing values are removed from
    the analysis. The results (present only that of $\mbox{SNP}_{41}, \mbox{SNP}_{42}$ and 
    $\mbox{SNP}_{50}$ for brevity) are given in the middle panel of Table~\ref{tab-2}.
    \item[A3]
    Same as A1, but only SNPs 36 to 50 are included in the analysis where 333 individuals do not
    have any missing values now. The results (restricted to present only that of $\mbox{SNP}_{41}, 
    \mbox{SNP}_{42}$ and $\mbox{SNP}_{50}$ for brevity) are given in the bottom panel of Table~\ref{tab-2}.
\end{itemize}

\begin{table}[!htbp]
	\centering
	\caption{Effects of $\mbox{SNP}_{41} (g_{41]}), \mbox{SNP}_{42} (g_{42}) $ and 
	$\mbox{SNP}_{50} (g_{50})$ on the phenotype in Simulation~1, obtained from logistic regression 
	and random forest with/without including the missing values.} \vspace{8pt}
	\begin{tabular}{|c|cccc|}
		\hline
		\multicolumn{5}{|c|}{Before missing data created; for all 1000 individuals $\times$ 100 SNPs}\\\hline
		\multicolumn{5}{|c|}{$\mbox{logit Pr}(y_i=1) =\beta_{0} + \beta_{1,1}g_{i,1}^{(1)}+ \beta_{1,2}g_{i,1}^{(2)}+\cdots+\beta_{100,1}g_{i,100}^{(1)}+\beta_{100,2}g_{i,100}^{(2)}$} \\ \hline
		&estimation&standard errors&p-values&MDA\\\hline
		(Intercept)&-3.148&0.645&1.07e-6***&\\
		$g_{41}^{(1)}$ & 2.979&0.518&9.00e-9***&\multirow{2}{*}{2.500e-2}\\
		$g_{41}^{(2)}$ &2.846&0.813&0.00463***&\\
		$g_{42}^{(1)}$ &2.561&0.494&2.12e-7***&\multirow{2}{*}{3.710e-2}\\
		$g_{42}^{(2)}$ &-2.594&0.902&0.004**&\\
		$g_{50}^{(1)}$ &-3.114&0.528&3.60e-9***&\multirow{2}{*}{8.044e-3}\\
		$g_{50}^{(2)}$ &-3.898&0.819&1.96e-6***&\\
		\hline \hline 
		\multicolumn{5}{|c|}{After missing data created; 114 individuals$\times$100 SNPs
		having no NAs}\\ \hline
		\multicolumn{5}{|c|}{$\mbox{logit Pr}(y_i=1) =\beta_{0} + \beta_{1,1}g_{i,1}^{(1)}+ \beta_{1,2}g_{i,1}^{(2)}+\cdots+\beta_{100,1}g_{i,100}^{(1)}+\beta_{100,2}g_{i,100}^{(2)}$} \\ \hline
		(Intercept) &-1.506e+2&1.277e+6&1&\\
		$g_{41}^{(1)}$ &4.265e+1&1.102e+6&1&\multirow{2}{*}{7.543e-03 }\\
        $g_{41}^{(2)}$ &-3.520e+2&2.302e+6&1&\\
        $g_{42}^{(1)}$ &1.306e+2&1.604e+6&1&\multirow{2}{*}{2.778e-02}\\
        $g_{42}^{(2)}$ &7.256e+2&8.109e+6&1&\\
        $g_{50}^{(1)}$ &-1.356e+2&1.120e+6&1&\multirow{2}{*}{2.673e-03}\\
        $g_{50}^{(2)}$ &-4.437e+2&4.847e+6&1&\\
		\hline \hline
		\multicolumn{5}{|c|}{After missing data created; 333 individuals$\times$15 SNPs having no NAs}\\\hline
		\multicolumn{5}{|c|}{$\mbox{logit Pr}(y_i=1) =\beta_{0}+\beta_{36,1}g_{i,36}^{(1)}+ \beta_{36,2}g_{i,36}^{(2)} +\cdots+\beta_{50,1}g_{i,50}^{(1)}+\beta_{50,2}g_{i,50}^{(2)}$}\\\hline
		(Intercept)&-3.579&0.618&6.83e-9***&\\
		$g_{41}^{(1)}$ &1.991&0.704&0.005**&\multirow{2}{*}{2.472e-02}\\
		$g_{41}^{(2)}$ &0.524&1.295&0.686&\\
		$g_{42}^{(1)}$ &2.831&0.778&0.00273***&\multirow{2}{*}{4.612e-02}\\
		$g_{42}^{(2)}$ &-0.862&1.772&0.627&\\
		$g_{50}^{(1)}$ &-2.782&0.655&2.16e-5***&\multirow{2}{*}{2.508e-02}\\
		$g_{50}^{(2)}$ &-1.812&1.156&0.117&\\
		\hline 
	\end{tabular}
	\label{tab-2}
\end{table}

Following conclusions can be drawn from Table~\ref{tab-2}:
\begin{itemize}
    \item[C1]
    In situations of no missing data and sparse phenotype-SNPs association, standard random forest and
    logistic regression analysis can correctly select the underpinning true SNPs (i.e. $g_{41}, g_{42}, g_{50}$)
    and their association effects, in that the MDA values for $g_{41}, g_{42}, g_{50}$ are the highest
    three, and $\hat{\beta}$ values $g_{41}^{(k)}, g_{42}^{(k)}, g_{50}^{(k)}$ ($k=1,2$) are all statistically
    significant (i.e. small $p$-values) and have the same signs as the true $\beta$ values. But the standard
    errors of $\hat{\beta}$'s are bit too large, possibly due to the high correlations among the 100 SNPs and
    using all the 100 SNPs for fitting the logistic model rather than being preceded by variable selection.
    \item[C2]
    When missing values having non-ignorable missingness mechanism are ignored and the number of candidate SNPs
    is large, random forest can still correctly select the underpinning true SNPs, but the logistic regression
    analysis cannot proceed properly (i.e standard errors of $\hat{\beta}$'s do not converge), due to the 
    co-linearity in the SNPs induced after too many individuals having missing values are excluded from analysis.
    \item[C3]
    When the number of candidate SNPs is not large, the adverse impact of ignoring missing values on the 
    analysis may not be as severe as seen in C2.
\end{itemize}

Next, we use our proposed method to analyze the phenotype-SNP associations, which proceeds as following:
\begin{itemize}
    \item[A4]
    Run the \textit{Ridge-EM $\rightarrow$ multiple imputations $\rightarrow$ complete data $\rightarrow$ RF
$\rightarrow$ variable selection $\rightarrow$ Ridge-EM} cycle for $\tau=10$ times, when the result has achieved
stabilization in that the SNPs selected do not change in two consecutive cycles. In the last cycle here, the
observed and imputed data together make up 6903 rows. Specifically, the number of imputations for each sample
individual having missing SNP values is set to be 3 or 9 or 10, respectively, if the number of missing
SNPs is 1, 2, or more than 2. Also there is no need to use ridge regression for this dataset. Thus we
set $\lambda=0$. The frequency of each SNP being selected by RF in those 10 cycles is displayed in Table~\ref{tab-3},
from which we see SNPs~9, 32, 39, 40, 41, 42, 43, 49, 50, 51 and 100, which cover $\mbox{SNP}_{41}, \mbox{SNP}_{42}$ and
$\mbox{SNP}_{50}$ used in the data generation process (\ref{eq-18}), in all cycles. We also see SNPs~1,5, 17 and 28 
are not selected at all.
\item[A5]
Fit the complete data, consisting of the 6903 rows of imputations or observations of the 100 SNPs plus their
missingness indicators and the phenotype, to the model system (\ref{eq-1}, \ref{eq-4}, \ref{eq-8}). 
The results are summarized in the top panel of Table~\ref{tab-4} where we present only the results concerning
$\mbox{SNP}_{41}, \mbox{SNP}_{42}$ and $\mbox{SNP}_{50}$.
\item[A6]
Do the same as A5 except the model system (\ref{eq-1}, \ref{eq-4}, \ref{eq-8}) includes
SNPs~9, 32, 39, 40, 41, 42, 43, 49, 50, 51 and 100 that are selected by RF (instead of including all the 100 SNPs). 
The results are given in the bottom panel of Table~\ref{tab-4}.
\end{itemize}

\begin{table}[!htbp]
	\centering
	\caption{Frequencies of each SNP being selected among the $\tau=10$ iterations in Simulation 1.} \vspace{8pt}
	\begin{tabular}{|cc|cc|}
		\hline
		Frequencies&SNP No.&Frequencies&SNP No.\\\hline
		\multirow{2}{*}{10}&9,32,39,40,41,42,&\multirow{2}{*}{5}&10,11,15,2,23,44,53,\\
		&43,49,50,51,100&&54,55,68,7,90\\\hline
		\multirow{2}{*}{9}&14,30,35,38,48,52,61,62,64,74,&\multirow{2}{*}{4}&\multirow{2}{*}{16,25,33,67,8,88}\\
		&78,80,83,84,93,94,95,96,98&&\\\hline
		8&13,20,27,29,31,46,59,65,77,81&3&34,4,56,6,72,85,86\\\hline
		\multirow{2}{*}{7}&12,21,36,37,63,70,&\multirow{2}{*}{2}&\multirow{2}{*}{18,19,57,58,71,73}\\
		&79,82,91,92,97,99&&\\\hline
		6&26,45,47,60,66,69,75,89&2&22,24,3,76,87\\
	\hline
    \end{tabular}
	\label{tab-3}
\end{table}

\begin{table}[!htbp]
	\centering
	\caption{Results of phenotype-SNP association analysis from the complete data including multiple imputations in Simulation 1.}
	\vspace{8pt}
	\begin{tabular}{|c|cccc|}
		\hline
		\multicolumn{5}{|c|}{Observed data plus imputations: 6903 individuals $\times$ 100 SNPs}\\\hline
		\multicolumn{5}{|c|}{$\mbox{logit Pr}(y_i=1) =\beta_{0} + \beta_{1,1}g_{i,1}^{(1)}+ \beta_{1,2}g_{i,1}^{(2)}+\cdots+\beta_{100,1}g_{i,100}^{(1)}+\beta_{100,2}g_{i,100}^{(2)}$}\\\hline
				&estimation&standard errors&p-values&Wald statistics\\\hline
		(Intercept)&-2.311&0.577&6.16e-5***&\\
		$g_{41}^{(1)}$ &1.743&0.436&6.27e-5***&\multirow{2}{*}{3.776e-07 ***}\\
		$g_{41}^{(2)}$ &-0.278&0.637&0.662&\\
		$g_{42}^{(1)}$ &2.910&0.431&1.39e-11***&\multirow{2}{*}{3.97e-12 ***}\\
		$g_{42}^{(2)}$ &0.600&0.782&0.443&\\
		$g_{50}^{(1)}$ &-2.972&0.489&1.24e-9***&\multirow{2}{*}{1.436e-09 ***}\\
		$g_{50}^{(2)}$ &-1.742&0.724&0.016*&\\
		\hline \hline
		\multicolumn{5}{|c|}{Include the RF selected SNPs 9,32,39,40,41,42,43,49,50,51,100 only in the analysis}\\\hline
		\multicolumn{5}{|c|}{$\mbox{logit Pr}(y_i=1) =\beta_{0} + \beta_{9,1}g_{i,9}^{(1)}+ \beta_{9,2}g_{i,9}^{(2)}+\cdots+\beta_{100,1}g_{i,100}^{(1)}+\beta_{100,2}g_{i,100}^{(2)}$}\\\hline
		(Intercept)&-1.889&0.277&9.36e-12***&\\
		$g_{9}^{(1)}$ &-0.467&0.185&0.0114*&\\
		$g_{9}^{(2)}$ &-0.024&0.288&0.935&\\
		$g_{32}^{(1)}$ &-0.191&0.191&0.317&\\
		$g_{32}^{(2)}$ &0.099&0.262&0.706&\\
		$g_{39}^{(1)}$  &0.355&0.247&0.150&\\
		$g_{39}^{(2)}$ &-0.653&0.477&0.171&\\
		$g_{40}^{(1)}$ &-0.281&0.318&0.377&\\
		$g_{40}^{(2)}$ & 0.080&0.501&0.873&\\
		$g_{41}^{(1)}$ &1.340&0.318&2.58e-5***&\multirow{2}{*}{1.84e-07 ***}\\
		$g_{41}^{(2)}$ &-0.049&0.482&0.920&\\
		$g_{42}^{(1)}$ &2.017&0.318&2.39e-10&\multirow{2}{*}{4.725e-11 ***}\\
		$g_{42}^{(2)}$ &0.385&0.624&0.537&\\
		$g_{43}^{(1)}$ &-0.004&0.277&0.988&\\
		$g_{43}^{(2)}$ &-0.161&0.376&0.669&\\
		$g_{49}^{(1)}$ &0.471&0.277&0.090&\\
		$g_{49}^{(2)}$ &0.682&0.449&0.129&\\
		$g_{50}^{(1)}$ &-2.110&0.341&6.392-10***&\multirow{2}{*}{4.829e-10 ***}\\
		$g_{50}^{(2)}$ &-1.252&0.532&0.0185*&\\
		$g_{51}^{(1)}$ &0.028&0.267&0.916&\\
		$g_{51}^{(2)}$ &0.122&0.412&0.768&\\
		$g_{100}^{(1)}$ &-0.069&0.183&0.709&\\
		$g_{100}^{(2)}$ &0.053&0.290&0.855&\\
		\hline
	\end{tabular}
	\label{tab-4}
\end{table}

From Table~\ref{tab-4} we can conclude that the RF in our method is able to select all SNPs underpinning
the true phenotype-SNPs associations, although it tends to over-select, i.e. some insignificant or
redundant SNPs may also be selected by RF. On the other hand, most of those insignificant or redundant
SNPs are found to have large $p$-values based on linear hypothesis testing in logistic regression analysis.
Overall, our method has capacity to identify and estimate the phenotype-SNPs associations effectively
and efficiently.

\subsection{Simulation study 2} \label{sec-3.3}
\begin{figure}[!htbp]	
	\includegraphics[width=6.0in]{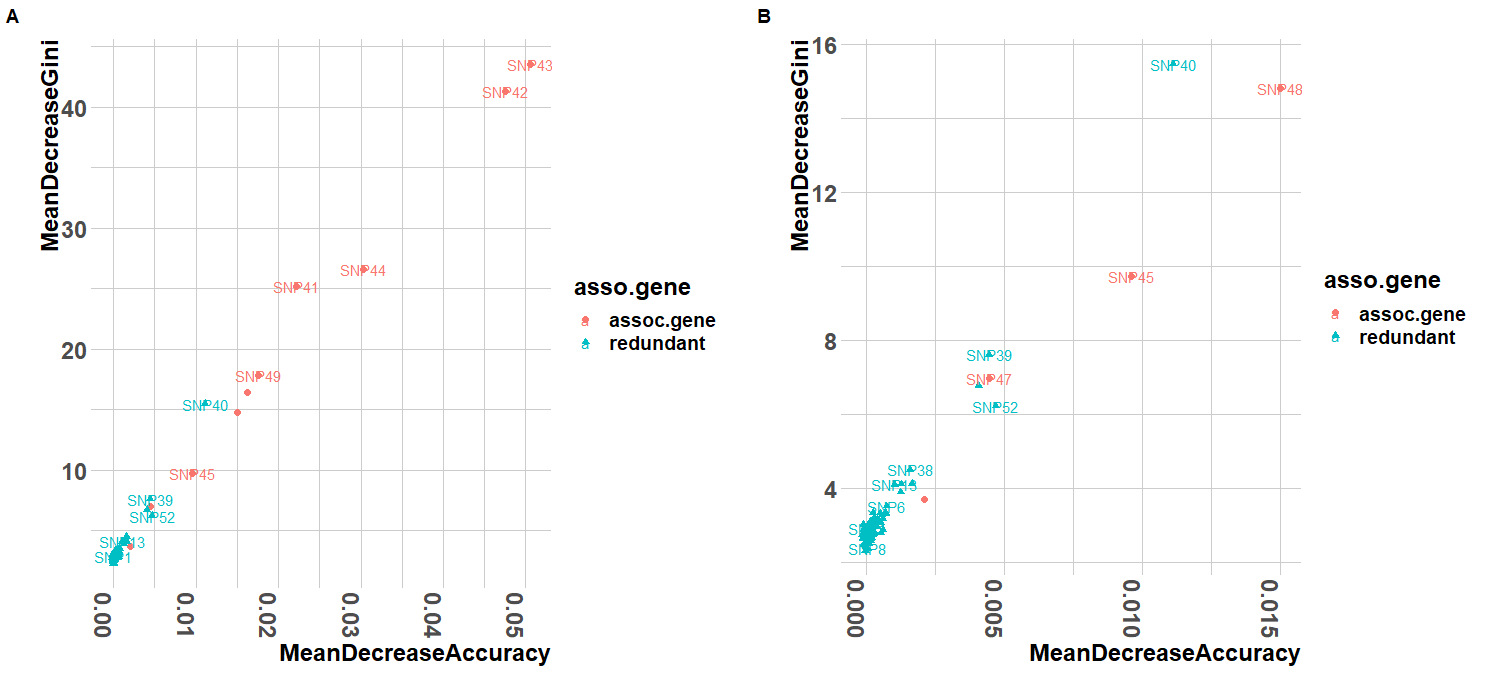}
	\centering
	\caption{Panel \textbf{A}: Scatter plot of Mean Decrease Gini (MDG) versus Mean Decrease Accuracy (MDA) values of the 100 SNPs 
	from the random forest analysis	on the generated data in Simulation~2. Panel \textbf{B} is an enlargement of the lower left
	corner of panel \textbf{A}. The red dots represent those SNPs that are used in the data generation process, while the green 
	dots are those not used.}
	\label{Fig3}
\end{figure}

Now we consider the situation where the phenotype response is associated with a sizeable number of SNPs
and there are missing values in both the phenotype and the SNPs. We first generate 1000 values of the
binary phenotype $y$ using the SNPs data matrix $G$ and the logistic regression model:
\begin{eqnarray}
\log\frac{\pi_i^*}{1-\pi_i^*}
&=&-2.2+1.2g_{i,41} +1.8g_{i,42}+1.6g_{i,43}+1.3g_{i,44}+1.7g_{i,45} \nonumber \\
& & -0.8g_{i,46}-g_{i,47}-0.9g_{i,48}-1.4g_{i,49}-g_{i,50}
	\label{eq-20}
\end{eqnarray}
where $\pi_i^*=\mbox{Pr}(y_{i}=1)$, $i=1,\cdots, 1000$. Note each SNP predictor in (\ref{eq-20}) is treated
as a numerical variable taking value 0, 1 or 2. If we encode $g_{i,j}$ by two dummy variables $g_{i,j}^{(1)}$
and $g_{i,j}^{(2)}$, equation (\ref{eq-20}) becomes
\begin{equation}
\log\frac{\pi_i^*}{1-\pi_i^*}
=-2.2+1.2g_{i,41}^{(1)}+2.4g_{i,41}^{(2)} +1.8g_{i,42}^{(1)}+3.6g_{i,42}^{(2)}+\cdots -g_{i,50}^{(1)}-2.0g_{i,50}^{(2)}.
	\label{eq-21}
\end{equation}
Hence (\ref{eq-21}) is equivalent to (\ref{eq-20}). For simplicity of presentation our analysis is based on
treating each SNP as a numerical variable. The specific values for the regression coefficient parameters in
(\ref{eq-20}) are determined by expecting the intercept parameter values result in 10\% base phenotype
prevalence (which is achieved since $e^{-2.2}/(1+e^{-2.2})\approx 0.1$), and expecting the $p$-values for
testing the significance of the other regression coefficient parameters be smaller than 0.05. To verify
the latter expectation, we fit a logistic regression model including all the 100 SNPs to the generated
$y$ data and $G$ matrix. The parameter estimation and testing results are presented in Table~\ref{Table1},
where only the results for $g_{41}$ to $g_{50}$, $g_{36}$ to $g_{40}$, and $g_{51}$ to $g_{55}$ are given
since the results for the other SNPs, which are not included in (\ref{eq-20}) or not significantly associated
with those in (\ref{eq-20}), are not relevant to the verification. A random forest is also fitted to the data, 
with the VIP measure MDG for each SNP variable computed and presented in Table~\ref{Table1}. Further, we plot 
the MDG values versus values of the other VIP measure MDA in Figure~\ref{Fig3}. Results from Table~\ref{Table1} 
and Figure~\ref{Fig3} confirm that the generated $y$ data can be regarded as being generated from model
(\ref{eq-20}) with strong statistical evidence in terms of the $p$-values. 

 
 \begin{table}[!htbp]
	\centering
	\caption{Results of logistic regression and random forest analysis on the generated data in Simulation 2.} \vspace{8pt}
	\begin{tabular}{|ccccc|}
		\hline
		&estimation&standard error&$p$-value&MDG\\ \hline
		SNP41 ($g_{41}$) &0.462&0.439&0.293&25.220\\
		SNP42 ($g_{42}$) &2.570&0.461&2.56e-8***&41.288\\
		SNP43 ($g_{43}$) &2.260&0.452&6.00e-7***&43.494\\
		SNP44 ($g_{44}$) &2.318&0.491&2.38e-6***&26.558\\
		SNP45 ($g_{45}$) &3.110&0.604&2.68e-7***&9.724\\
		SNP46 ($g_{46}$) &-2.083&0.597&0.000481***&3.704\\
		SNP47 ($g_{47}$) &-0.962&0.486&0.05*&6.978\\
		SNP48 ($g_{48}$) &-2.235&0.554&5.41e-5***&14.819\\
		SNP49 ($g_{49}$) &-1.241&0.468&0.008**&17.798\\
		SNP50 ($g_{50}$) &-2.137&0.440&1.19e-6***&16.418\\
		\hline
		SNP36 ($g_{36}$) &0.519&0.440&0.238&3.274\\
		SNP37 ($g_{37}$) &0.008&0.422&0.984&4.114\\
		SNP38 ($g_{38}$) &-0.154&0.437&0.724&4.506\\
		SNP39 ($g_{39}$) &-0.525&0.462&0.256&7.618\\
		SNP40 ($g_{40}$) &0.629&0.418&0.133&15.478\\
		SNP51 ($g_{51}$) &0.404&0.409&0.322&6.774\\
		SNP52 ($g_{52}$) &-0.286&0.492&0.561&6.223\\
		SNP53 ($g_{53}$) &0.553&0.479&0.248&4.127\\
		SNP54 ($g_{54}$) &-0.600&0.436&0.169&3.895\\
		SNP55 ($g_{55}$) &-1.116&0.451&0.013*&3.334\\
		\hline
	\end{tabular}
\label{Table1}
\end{table}

\begin{figure}[!htbp]	
	\includegraphics[width=6.0in]{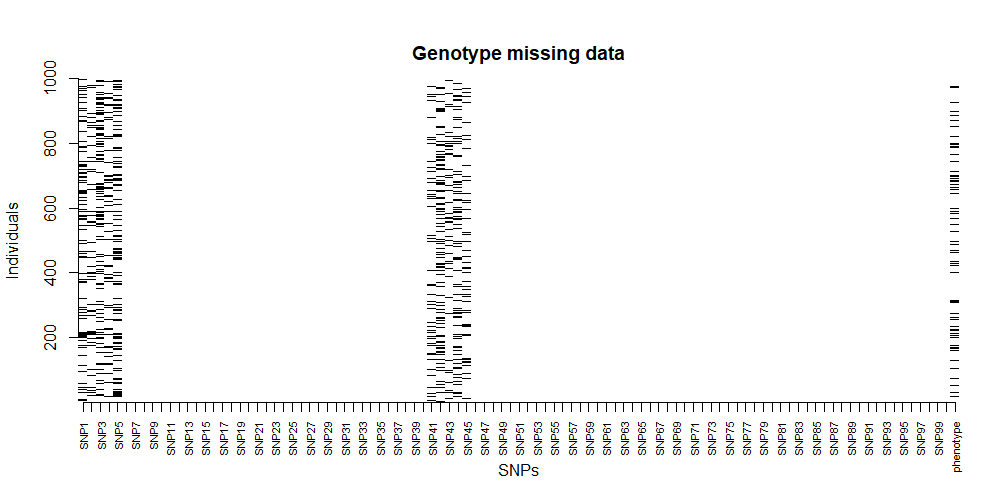}
	\centering
	\caption{Missingness patterns for SNPs and the phenotype in Simulation 2. Black bar at a (SNP, individual) location 
	indicates the corresponding SNP value is missing for the referenced individual. Blank bar indicates it is observed.}
	\label{Fig2-sim2}
\end{figure}

We now turn some observations in $G$ according to the missingness mechanism (\ref{eq-19}), and turn some $y$ values
to missing as well according to the missingness mechanism
\begin{equation}
\frac{\mbox{Pr}(r(y_i)=1)}{\mbox{Pr}(r(y_i)=0)}=-2.0+1.2y_i, \quad i=1,...,1000,
	\label{eq-22}
\end{equation}
where $r(y_i)=1$ or 0 depending on whether or not $y_i$ is missing. It is found that the generated $y$ data contain
205 missing values. Proportions of missing values in the observations of
$\mbox{SNP}_1, \cdots, \mbox{SNP}_5$ and $\mbox{SNP}_{41},\cdots,\mbox{SNP}_{45}$ are given in Table~\ref{tab-6}, and
their missingness patterns together with that of $y$ are summarized in Figure~\ref{Fig2-sim2}.
It can be found that the total number of missing values in $\mbox{SNP}_1,\cdots, \mbox{SNP}_5, \mbox{SNP}_{41}, \cdots, 
\mbox{SNP}_{45}$ in Simulation~2 is 1974, and the numbers of 
individuals having 0, 1, 2, 3, 4, 5, 6, 7 and 8 missing values are 113, 241, 271, 200, 108, 45, 16, 5 and 1, respectively.

\begin{table}[!htbp]
		\centering
	\caption{Parameter values used in the SNPs missingness mechanism model and the resultant proportions of missing
	SNP values in Simulation 2.} \vspace{6pt}
	\begin{tabular}{|c|cccccccccc|} \hline
		$\mbox{SNP}_j$ & $g_1$ & $g_2$ & $g_3$ & $g_4$ & $g_5$ & $g_{41}$ & $g_{42}$ & $g_{43}$ & $g_{44}$ & $g_{45}$ \\
		\hline
		$\xi_{j,0}^*$ & $-2.0$ & $-2.0$ & $-2.0$ & $-2.0$ & $-2.0$ & $-2.0$ & $-2.0$ & $-2.0$ & $-2.0$ & $-2.0$  \\
		$\xi_{j,1}^*$ & $1.1$ & $0.4$ & $1.1$ & $0.4$ & $1.1$ & $0.4$ & $1.1$ & $0.4$ & $1.1$ & $0.4$  \\
		\%missing & .236 & .159 & .259 & .141 & .253 & .153 & .245 & .145 & .236 & .147 \\ \hline
	\end{tabular}
\label{tab-6}
\end{table}

Next we apply our proposed method to analyze the phenotype-SNP associations using the latest generated $y$ and
$G$ data that contain missing values with non-ignorable missingness mechanism. The procedure and results are
described in three parts as following.

\begin{figure}[!htbp]	
	\includegraphics[width=6in]{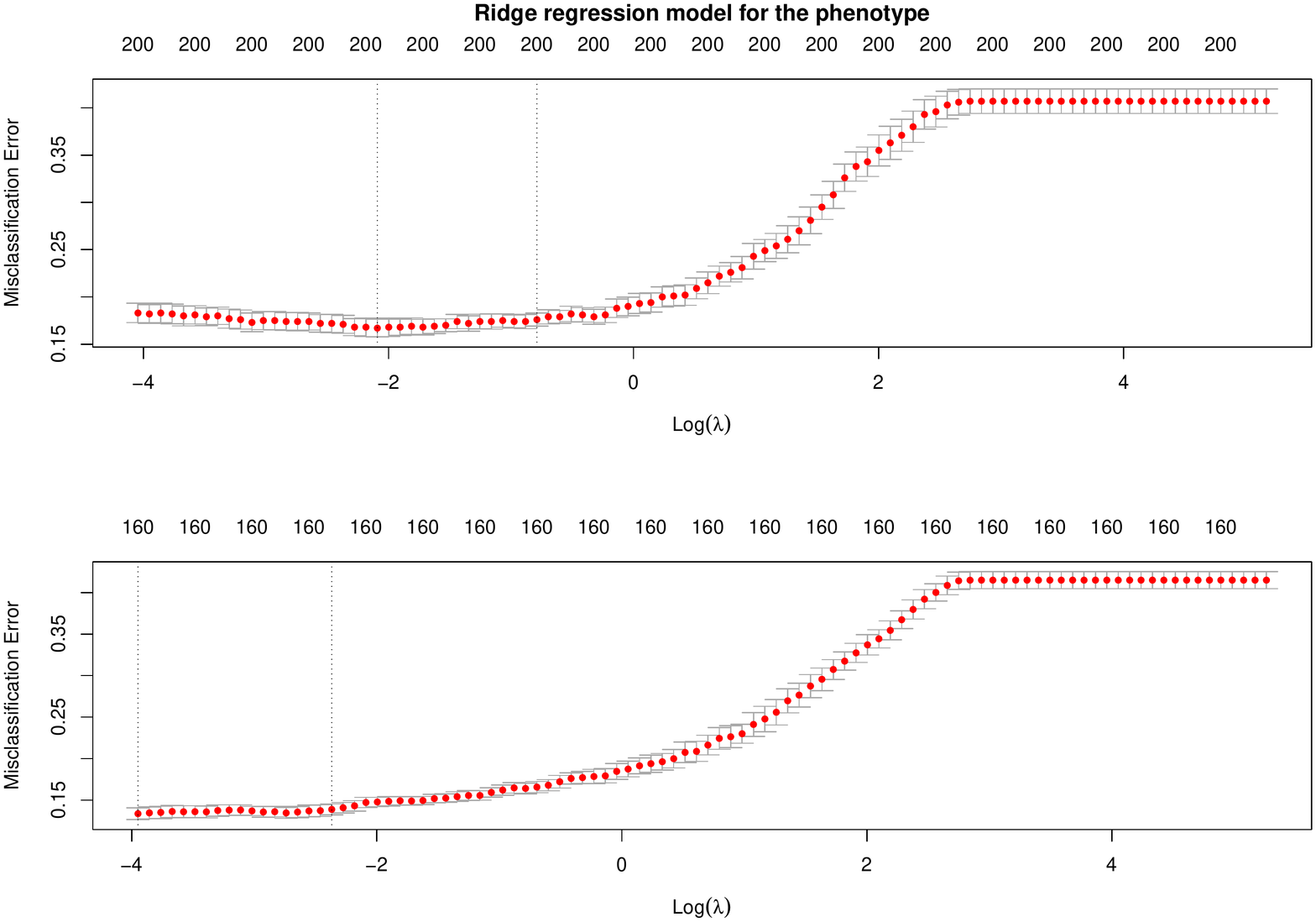}
	\centering
	\caption{The \textit{top} plot shows the mis-classification error (ME) trajectory as the tuning parameter $\lambda$ changes 
	in the first run of Ridge-EM. The ME is minimized at $\log \lambda=-2.091$. The \textit{bottom} plot is obtained from the 
	last run of Ridge-EM. The ME is minimized at $\log\lambda=-3.949$. There are two vertical lines in each plot. The left line refers to $\log\lambda$ that gives the minimum ME and the right line refers to the largest value of $\log\lambda$ that 
	is within 1 standard error of the minimum ME.}
	\label{Fig2-cv}
    \end{figure}

\begin{figure}[!htbp]	
	\includegraphics[width=6in]{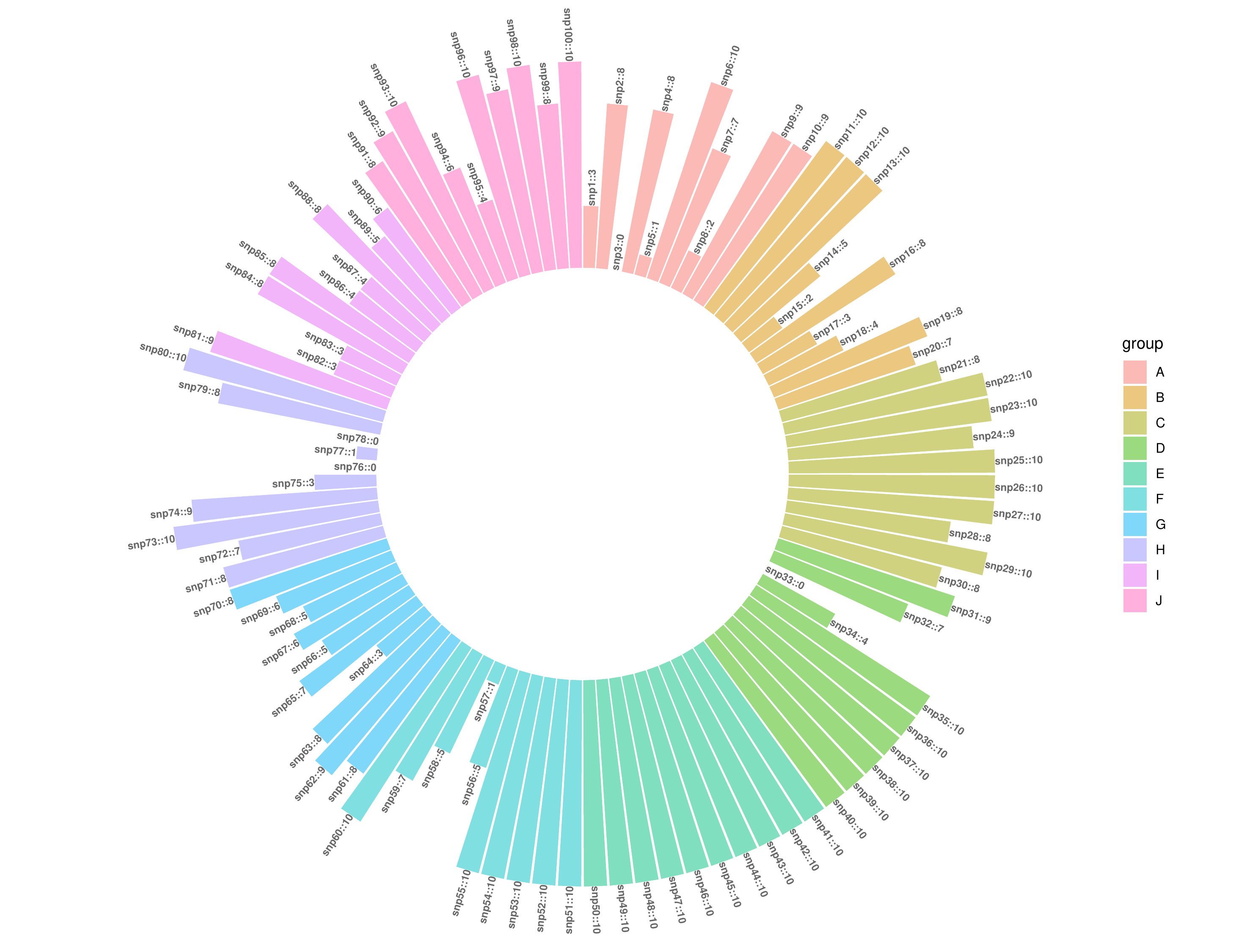}
	\centering
	\caption{Circular bar plot for frequencies among $100$ SNPs. The top of each bar shows the related SNP name and frequency.
	The tuning parameter $\lambda$ involved in the computation is determined by cross-validation.}
	\label{Fig4}
\end{figure}

\paragraph{Part 1.} We run the \textit{Ridge-EM $\rightarrow$ multiple imputations $\rightarrow$ complete data $\rightarrow$ RF
$\rightarrow$ variable selection $\rightarrow$ Ridge-EM} cycle for $\tau=10$ times, where the ridge tuning parameter $\lambda$
is chosen to minimize the predictive mis-classification error by cross-validation. The cross-validation results are shown
in Figure~\ref{Fig2-cv}, from the top panel of which the best $\lambda$ value is found to be $\lambda=e^{-2.091}=0.124$ in
the first run of the cycle; and the best $\lambda=e^{-3.949}=0.019$ (cf. bottom panel of Figure~\ref{Fig2-cv}) is found
in the tenth run of Ridge-EM when SNPs variable selection is completed. Frequencies of the 100 SNPs being respectively
selected by RF for equation (\ref{eq-1}) out of the 10 cycles are summarized by a circular bar plot displayed in 
Figure~\ref{Fig4}. It is shown from Figure~\ref{Fig4} that 38 SNPs, including $g_{41}\sim g_{50}$ that are used in
the data generation and their highly correlated ones (i.e. $g_{35}\sim g_{40}$ and $g_{51}\sim g_{55}$), have their frequencies
10 out of 10. There are 26 SNPs having their frequencies $\leq 5$. We include only the former 38 SNPs into the model
system (\ref{eq-1}, \ref{eq-4}, \ref{eq-8}) for the phenotype-SNP association analysis, which gives the results very
similar to those used in Table~\ref{Table1}; thus not presented here for brevity of presentation.

We also test the missingness mechanism involved in the above association analysis, with the results provided in 
Table~\ref{Table2}. Note that the term ``related SNPs having missing values" in Table~\ref{Table2} refers to those
SNPs and phenotype $y$ that have missing values and appear in the missingness mechanism model (\ref{eq-8}). 
Results in Table~\ref{Table2} show that the missingness mechanisms in SNPs 41 to 45 and $y$ are non-ignorable
with strong statistical evidences. As an example, Table~\ref{Table2} shows that the missingness of SNP41 is 
significantly related to SNP45, thus to SNP41 as well because the correlation coefficient between SNP45 and
SNP41 is $0.8^4=0.41$ which is a sizeable number. Although the missingness mechanisms for SNPs 1, 2, 3, 4 and 5
are also found to be non-ignorable from Table~\ref{Table2}, they are not of our concern because SNPs 1 to 5
are not found to be significantly associated with the phenotype (cf. Figure~\ref{Fig4}).

\begin{figure}[htbp]
	\includegraphics[width=6.5in]{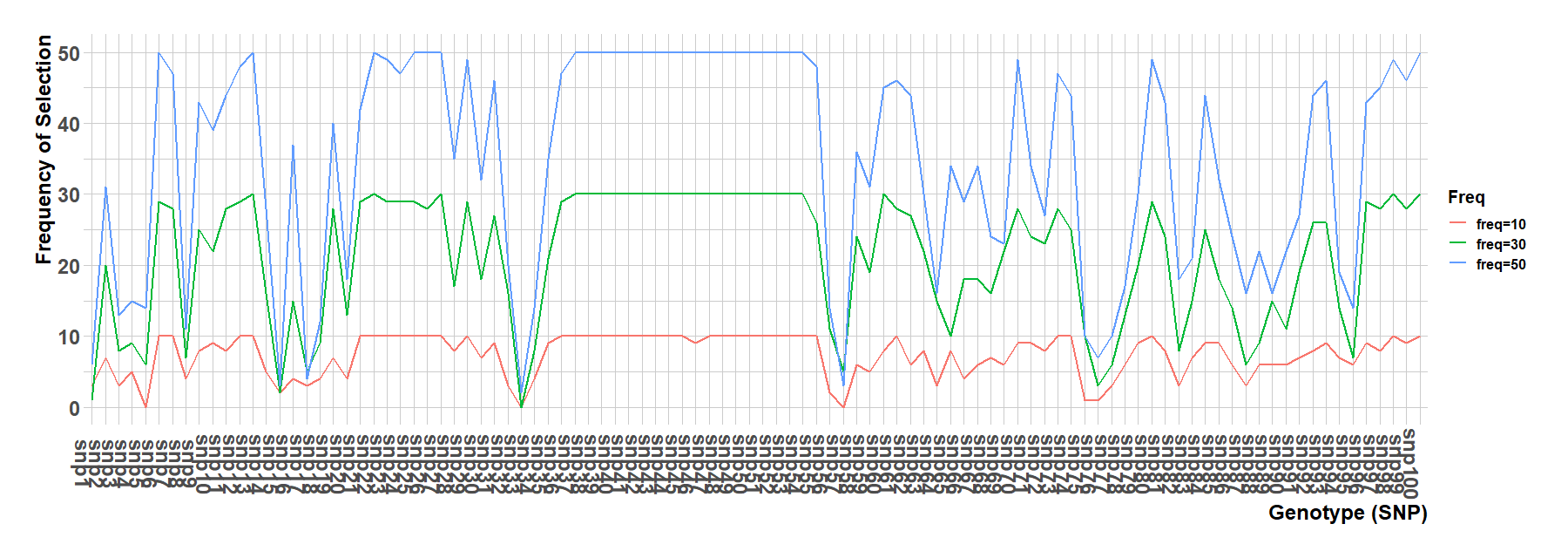}
	\centering
	\caption{Frequencies among 100 SNPs under different iterations. Blue line shows the frequencies result after 50 iterations; green line shows the frequencies result after 30 iterations and red line shows the frequencies result after 10 iterations.}
	\label{Fig7}
\end{figure}

\paragraph{Part 2.} We perform two further runs of the \textit{Ridge-EM $\rightarrow$ multiple imputations 
$\rightarrow$ complete data $\rightarrow$ RF $\rightarrow$ variable selection $\rightarrow$ Ridge-EM} cycle for 
$\tau=30$ and 50 times, respectively. The results on frequencies of each SNP in (\ref{eq-1}) being selected from 
the RF in each run are displayed in Figure~\ref{Fig7}. It can be found from Figure~\ref{Fig7} that number of
SNPs being selected in each cycle run is 38 when $\tau=10$; 24 when $\tau=30$; and 25 when $\tau=50$. This shows
that our method has strong capacity to identify all those SNPs that have significant associations with the
phenotype, but also has a tendency of over-selection.

 \begin{table}[htbp]
	\centering
	\caption{Results on the missingness mechanisms of those SNPs (or $y$) having missing values in Simulation~2} \vspace{8pt}
	\begin{tabular}{|c|c|c|}
		\hline
		Missingness & Related SNPs having missing values that & Wald test 
		\\
		indicators & appear in the missingness mechanism model (\ref{eq-8}) & $p$-value \\
		\hline
		$r$(SNP1) &2, 3, 4, 5&4.654e-8***
		\\
		$r$(SNP2)&3, 4, 44&0.00103**
		\\
		$r$(SNP3) &1, 2, 3, 4, 5&8.543e-13***
		\\
		$r$(SNP4) &2, 43&0.07721 $\cdot$
		\\
		$r$(SNP5) &3, 4, 5&8.142e-11***
		\\
		$r$(SNP41) &45&0.01916*
		\\
		$r$(SNP42)& 41, 42, 43, 44&0.005488***
		\\
		$r$(SNP43) &4, 43&0.02899
		\\
		$r$(SNP44) &41, 42, 43, 44, 45&4.209e-5***
		\\
		$r$(SNP45) &42, 44&0.005264**
		\\
	    $r$(phenotype) &41, 42, $y$ & 3.3e-7
	    \\
		\hline
	\end{tabular}
	\label{Table2}
\end{table}

\begin{figure}[!htbp]	
	\includegraphics[width=6in]{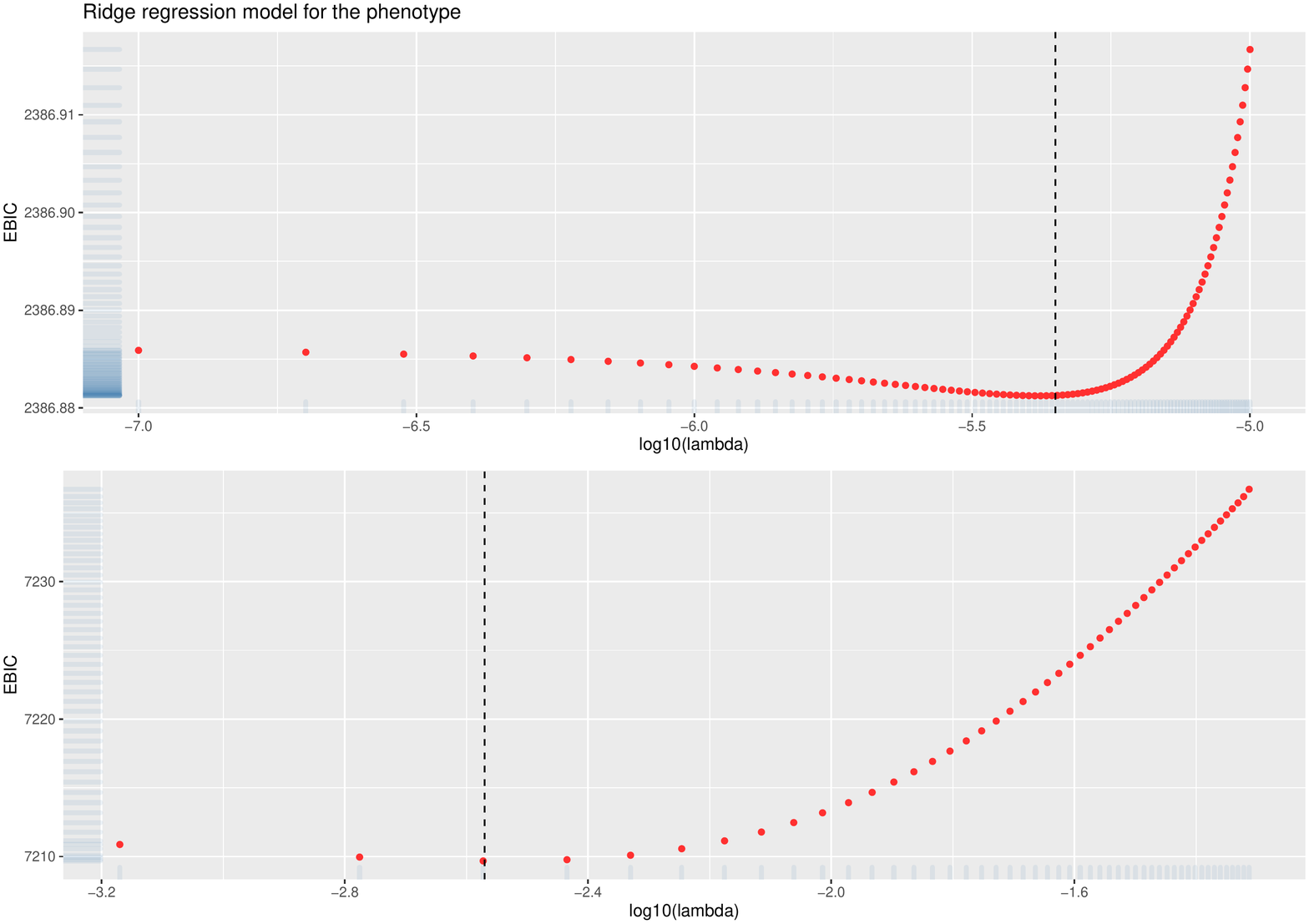}
	\centering
	\caption{The \textit{top} plot shows the EBIC trajectory as the tuning parameter $\lambda$ changes 
	in the first run of Ridge-EM in Simulation~2. The EBIC is minimized at $\log_{10}\lambda=-5.38$. The \textit{bottom} plot is 
	obtained from the last run of Ridge-EM. The EBIC is minimized at $\log_{10}\lambda=-2.57$.}
	\label{fig8-ebic}
    \end{figure}

\begin{figure}[t] 
	\includegraphics[width=6in]{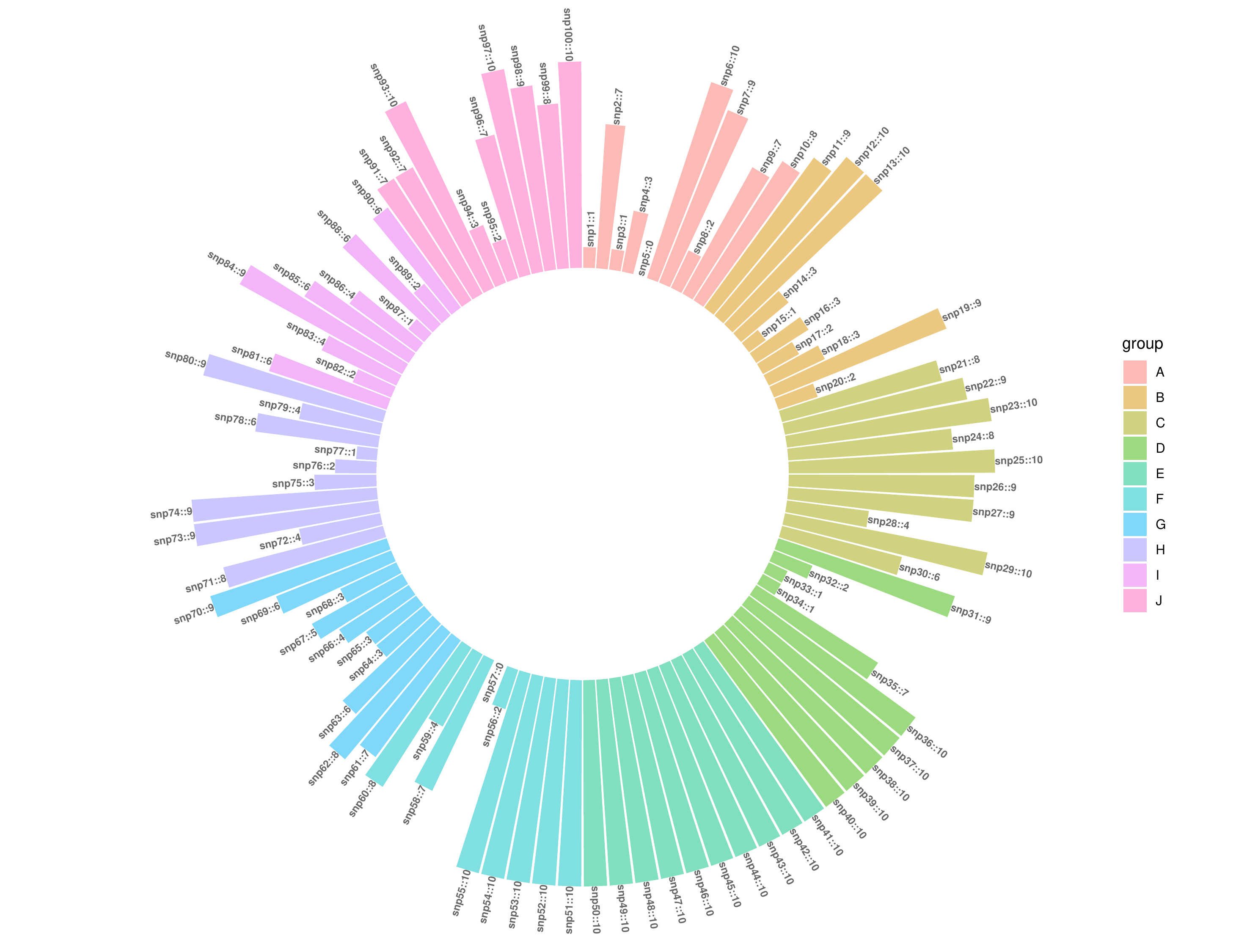}
	\centering
	\caption{Circular bar plot for frequencies among $100$ SNPs. The top of each bar shows the related SNP name and frequency.
	The tuning parameter $\lambda$ involved in the computation is determined by EBIC.}
	\label{Fig9}
\end{figure}

\paragraph{Part 3.} The same as Part 1 except that EBIC is used to tune the ridge parameter $\lambda$. It is found that
the best $\lambda$ value is $4.2\times 10^{-6}$ in the first run of the cycle, and is 0.003 in the Ridge-EM after SNP
variable selection is completed. The EBIC-based $\lambda$ tuning results are shown in Figure~\ref{fig8-ebic}.
Results on the frequencies of each SNP being selected from the RF for (\ref{eq-1}) in $\tau=10$ runs of the cycle
are summarized by the circular bar plot in Figure~\ref{Fig9}. From Figure~\ref{Fig9} we see 29 SNPs, including
SNPs 36 to 55, have full frequency 10; and 35 SNPs have their frequencies $\leq 5$. Hence, tuning $\lambda$ 
by EBIC rather than by cross-validation results in less SNPs being selected and more SNPs being removed.

\section{Real data application} \label{sec-4}
\begin{figure}[!htbp]	
	\includegraphics[width=6in]{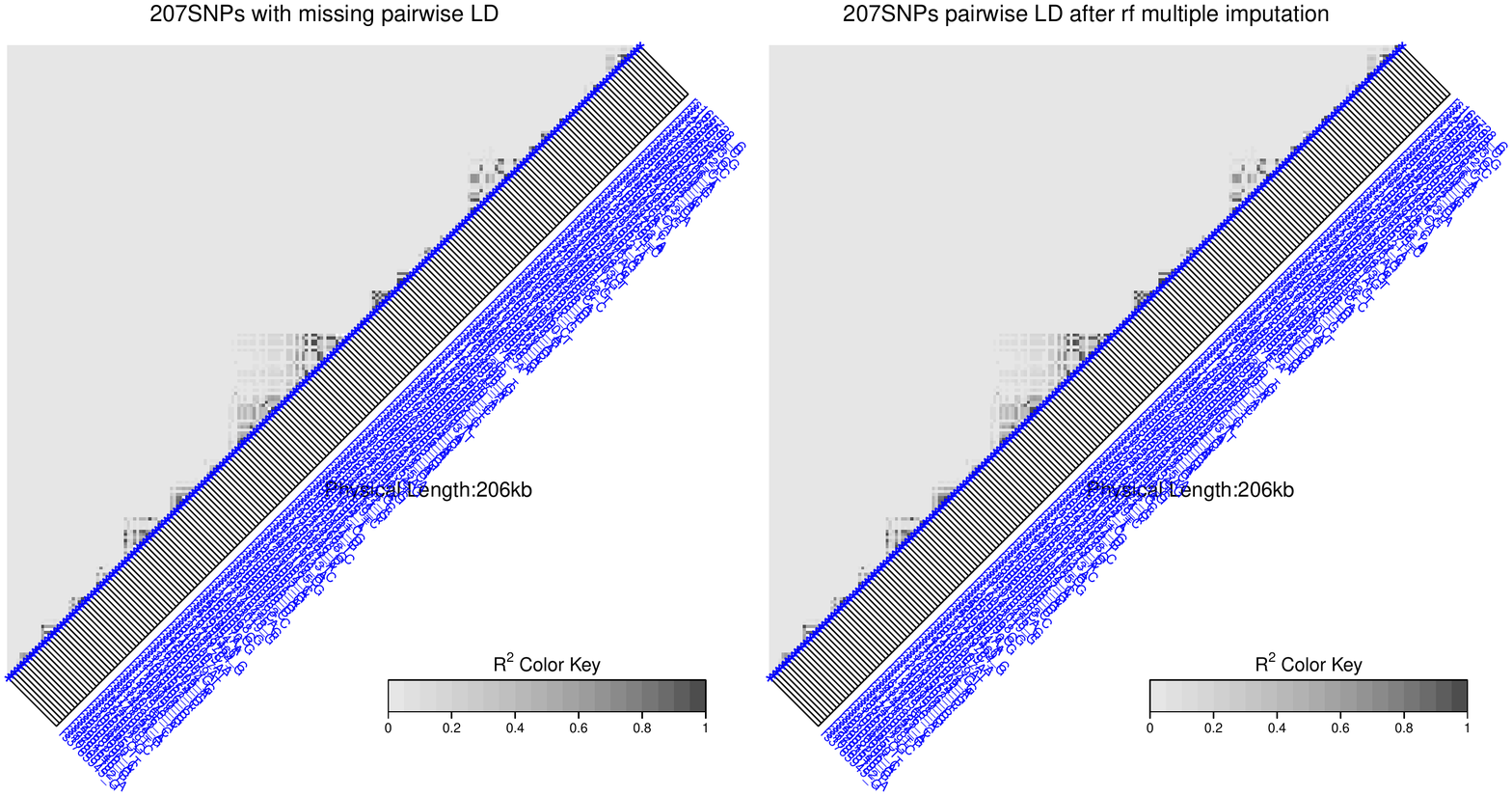}
	\centering
	\caption{\textit{Left}: LD pattern of the 207 SNPs where the missing values are imputed with those having
	the highest frequencies. \textit{Right}: LD pattern of the 207 SNPs where the missing values are replaced with
	the proposed multiple imputation generations.}
	\label{Fig10}
\end{figure}

Now we apply our proposed method to analyze the breast cancer data set introduced at the beginning of Section~\ref{sec-2},
which comprises 123,372 observations of 207 SNPs on 596 individuals (354 breast cancer patients and 242 matched controls).
There are 1,724 missing values in the data, distributions of which over the SNPs and the individuals are summarized in
Table~\ref{Table3} and Table~\ref{Table5}, respectively.
The pairwise LD structure of the 207 SNPs is displayed in Figure~\ref{Fig10} where single imputation for missing values
is used in the left plot and multiple imputations are used in the right one. Figure~\ref{Fig10} shows that, as the distance 
between any two SNPs increases, the dependence between them would decrease. On the other hand, there exists strong 
dependence between adjacent SNPs.

\begin{table}[!htbp]
	\centering
	\caption{Distribution of the range of missing values over SNPs (e.g. There are 81 SNPs each having $m$
	missing values with $1\leq m\leq 9$).} \vspace{8pt}
	\begin{tabular}{c|ccccc}
	\hline
	Range of numbers &\multirow{2}{*}{0}&\multirow{2}{*}{1-9}&\multirow{2}{*}{10-19}&\multirow{2}{*}{20-29}&\multirow{2}{*}{30-32}\\
	of missing values &&&&&\\\hline
	Number of SNPs &50 & 81 & 44 & 30 &2\\
		\hline
\end{tabular}
\label{Table3}
\end{table}

\begin{table}[!htbp]
	\centering
	\caption{Distribution of the range of missing values over individuals (e.g. There are 132 individuals
	each having 2 missing values).} \vspace{8pt}
	\begin{tabular}{c|cccccccccc}
		\hline
Range of numbers & \multirow{2}{*}{0} & \multirow{2}{*}{1} & \multirow{2}{*}{2} & \multirow{2}{*}{3} & \multirow{2}{*}{4} 
& \multirow{2}{*}{5} & \multirow{2}{*}{6} & \multirow{2}{*}{7} & \multirow{2}{*}{8} &\multirow{2}{*}{9} \\
	of missing values & & & & & & & & &\\\hline
Number of individuals & 83 & 107 & 132 & 79 & 59 & 44 & 31 & 30 & 15 & 16\\
		\hline
	\end{tabular}
	\label{Table5}
\end{table}
 
\begin{figure}[t] 
	\includegraphics[width=6in]{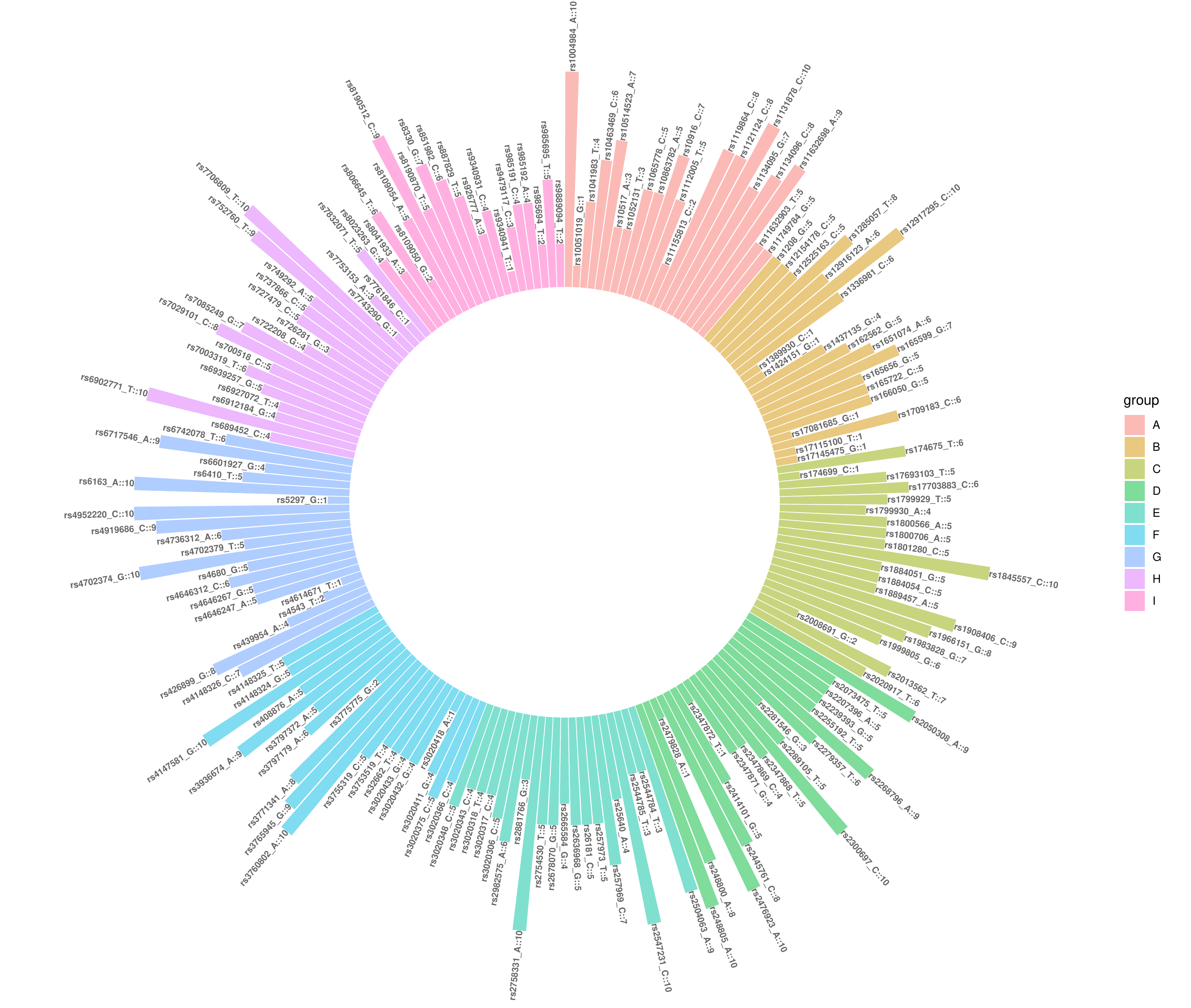}
	\centering
	\caption{Circular frequency map for the 207 SNPs of the breast cancer data. The top of each bar shows the related SNP name 
	and frequency. Twenty-eight SNPs of the 207 have 0 frequency, thus are not shown in the map.}
	\label{Fig11}
\end{figure}

The \textit{Ridge-EM $\rightarrow$ multiple imputations $\rightarrow$ complete data $\rightarrow$ RF
$\rightarrow$ variable selection $\rightarrow$ Ridge-EM} cycle is run $\tau=10$ times, resulting in 16 SNPs
being selected in each run, which can be determined from the circular frequency map in Figure~\ref{Fig11} and also
are listed in Table~\ref{Table6}. It is found that 11 of the 16 selected SNPs, which are displayed in parentheses 
in Tabele~\ref{Table6}, contain missing values. In this analysis, each SNP variable is coded with two dummy variables
as in Simulation~1. Since many of the SNPs having missing values have just a few missing values (cf. Table~\ref{Table3}),
we used the bias-reduction estimation method of \citeA{firth1993bias} and the corresponding \texttt{R} package
\texttt{brglm} \cite{kosmidis2020package} in each fitting of (\ref{eq-1},\ref{eq-4},\ref{eq-8}). Also there is no need
to use ridge regression for this particular data set. In other words, using $\lambda=0$ is sufficient for having
convergent parameter estimates.

We now use the breast cancer data, together with the multiple imputations of the missing values, to fit the logistic
model (\ref{eq-1}) where only the 16 selected SNPs are included as covariates. Results on the parameter estimation 
in model (\ref{eq-1}) are summarized in the left panel of Table~\ref{Table6}. For comparison, we fit the same
logistic model to the same breast cancer data but the involved missing values are excluded from fitting. 
The results are shown in the right panel in Table~\ref{Table6}.
Comparing the left panel with the right one in Table~\ref{Table6} we see \texttt{rs2547231C} is significantly associated 
with the breast cancer risk in both panels; \texttt{rs1004984A} is more significant in the left panel than in the 
right panel; \texttt{rs1845557C} is significant at level 0.1 in the left panel but not significant in the right panel; 
and \texttt{rs6902771T} is significant in the left panel but not in the right one. The other SNPs do not seem to be 
significant in either the panel. Therefore, by using multiple imputations to account for the effect of missing values 
we are able to find some significant SNPs that can not be found by using only the observed data to fit the model. 
There should be more analysis on whether any redundant SNPs can be further removed from the model (\ref{eq-1}) for 
the breast cancer data. But it would be beyond the scope of this research.

To see whether the missingness mechanism can be ignored for each of the 11 SNPs listed in Table~\ref{Table6} and
having missing values, we fit model (\ref{eq-8}) for these 11 SNPs as part of the whole process. The results are
summarized in Table~\ref{Table11}, from which we see all the 11 SNPs, except \texttt{rs1131878C} and \texttt{rs2547231C},
have non-ignorable missingness mechanism. Here we say a SNP \texttt{rsxxx} has non-ignorable missingness mechanism
if the missingness indicator $r$(\texttt{rsxxx}) is significantly associated with any SNPs having missing values
or with missingness indicators of any other SNPs.

\begin{table}[!htbp]
	\centering
	\caption{Results comparison for the logistic regression analysis of the phenotype-SNPs associations. 
	\textit{Left} panel is based on the proposed multiple imputation and random forest variable selection method.
	\textit{Right} panel is obtained from using only the observed data and ignoring the missing data. 
	SNPs shown in parentheses are those having missing values.} 
	\vspace{8pt}
	\label{Table6}
	\begin{tabular}{|c|ccc|ccc|}
		\hline
		&\multicolumn{3}{|c|}{\textbf{MI-RF based}}&\multicolumn{3}{|c|}{\textbf{Observed data based}}\\\hline
		Coefficients&Estimate &Std.Err & Pr($>|z|$)&Estimate &Std.Err&Pr($>|z|$)\\\hline
		Intercept &-0.165&0.515&0.748&0.143&0.539&0.791\\
		$(\mbox{rs1004984}_{A1})$&0.393&0.196& \textbf{0.045} &0.365&0.206& \textbf{0.077}\\
		$(\mbox{rs1004984}_{A2})$&0.200&0.279&0.473&0.252&0.292&0.387\\
		$(\mbox{rs1131878}_{C1})$&0.275&0.192&0.152&0.287&0.201&0.154\\
		$(\mbox{rs1131878}_{C2})$&-0.034&0.326&0.916&0.052&0.345&0.881\\
		$(\mbox{rs12917295}_{C1})$&-0.242&0.197&0.219&-0.200&0.208&0.335\\
		$(\mbox{rs12917295}_{C2})$&-0.305&0.270&0.259&-0.271&0.282&0.337\\
		$(\mbox{rs1845557}_{C1})$&0.050&0.191&0.792&0.036&0.202&0.857\\
		$(\mbox{rs1845557}_{C2})$&-0.516&0.304& \textbf{0.09}&-0.503&0.318&0.113\\
		$\mbox{rs2300697}_{C1}$&0.268&0.217&0.216&0.177&0.226&0.433\\
		$\mbox{rs2300697}_{C2}$&0.072&0.312&0.817&-0.033&0.325&0.919\\
		$\mbox{rs2476923}_{A1}$&-0.049&0.209&0.816&0.013&0.217&0.951\\
		$\mbox{rs2476923}_{A2}$&-0.324&0.260&0.213&-0.226&0.275&0.412\\
		$(\mbox{rs248805}_{A1}$)&0.130&0.287&0.651&0.112&0.302&0.711\\
		$(\mbox{rs248805}_{A2})$&0.502&0.376&0.182&0.357&0.399&0.371\\
		$(\mbox{rs2547231}_{C1})$&-0.726&0.196& \textbf{0.0002}&-0.683&0.204& \textbf{0.0008}\\
		$(\mbox{rs2547231}_{C2})$&0.418&0.724&0.564&0.041&0.751&0.956\\
		$(\mbox{rs2758331}_{A1})$&-0.345&0.219&0.115&-0.341&0.232&0.143\\
		$(\mbox{rs2758331}_{A2})$&-0.381&0.250&0.128&-0.430&0.263&0.102\\
		$\mbox{rs3760802}_{A1}$&0.075&0.213&0.723&-0.045&0.225&0.843\\
		$\mbox{rs3760802}_{A2}$&0.300&0.252&0.234&0.115&0.263&0.662\\
		$(\mbox{rs4147581}_{G1})$&0.285&0.211&0.177&0.328&0.220&0.135\\
		$(\mbox{rs4147581}_{G2})$&0.257&0.251&0.305&0.146&0.261&0.577\\
		$(\mbox{rs4702374}_{G1})$&0.491&0.774&0.526&-0.220&0.898&0.806\\		
		$(\mbox{rs4702374}_{G2})$&0.060&1.032&0.954&-0.741&1.196&0.536\\
		$\mbox{rs4952220}_{C1}$&0.175&0.217&0.421&0.156&0.226&0.491\\
		$\mbox{rs4952220}_{C2}$&0.288&0.310&0.354&0.233&0.324&0.473\\
		$(\mbox{rs6163}_{A1})$&-0.131&0.194&0.500&-0.067&0.205&0.745\\
		$(\mbox{rs6163}_{A2})$&-0.359&0.268&0.181&-0.365&0.277&0.188\\
		$\mbox{rs6902771}_{T1}$&0.434&0.207& \textbf{0.036} &0.307&0.218&0.158\\
		$\mbox{rs6902771}_{T2}$&0.031&0.253&0.902&-0.028&0.264&0.916\\
		$(\mbox{rs7706809}_{T1})$&-0.498&0.753&0.509&0.097&0.873&0.912\\
		$(\mbox{rs7706809}_{T2})$&0.846&1.026&0.409&1.442&1.185&0.224\\\hline
		&\multicolumn{3}{|c|}{\textbf{AIC}: 654.74}&\multicolumn{3}{|c|}{\textbf{AIC}: 737.13  }\\
		\hline
	\end{tabular}
\end{table}

\begin{table}[!htbp]
		\centering
		\caption{Missingness mechanism test for the 11 selected SNPs having NA values. 
		SNPs shown in parentheses are those having missing values. As an example, $r(\mbox{rs123}_{A})$
		means the missingness indicator of $\mbox{rs123}_{A}$.}
		\vspace{8pt}
		\label{Table11}
		\begin{tabular}{|c|c|c|c|c|}
			\hline
			Coefficient & Estimate & Std.Error& Z value& Pr($>|z|$)\\\hline
			\multicolumn{5}{|c|}{\textbf{Testing the missingness mechanism of $rs1131878_{C}$}}\\\hline
			
			Intercept&-6.659&1.472&-4.525& \textbf{6.05e-6}\\
			$(rs3765945_{G1})$&-0.466&1.716&-2.272&0.786\\
			$(rs3765945_{G2})$&1.724&1.473&1.170&0.242\\
			$rs3797179_{A1}$&1.157&1.615&0.716&0.474\\
			$rs3797179_{A2}$&4.236&1.463&2.895& \textbf{0.004}\\
			\hline
			\multicolumn{5}{|c|}{\textbf{Testing the missingness mechanism of $rs2547231_{C}$}}\\\hline
			Intercept&-5.900&1.420&-4.155& \textbf{3.26e-5}\\
			$rs6902771_{T1}$&-0.472&2.007&-0.235& 0.814\\
			$rs6902771_{T2}$&1.505&1.642&0.917& 0.359\\
			\hline      
			\multicolumn{5}{|c|}{\textbf{Testing the missingness mechanism of $rs1845557_{C}$}}\\\hline
			Intercept &-5.822&1.149&-5.068& \textbf{4.02e-7}\\
			$r(\mbox{rs4148326}_{C})$&1.268&1.366&0.928& 0.353\\
			$(rs1131878_{C1})$&0.261&0.857&0.305&0.760\\
			$(rs1131878_{C2})$&1.236&1.022&1.210&0.226\\
			$(rs1285057_{T1})$&-3.355&0.905&-0.392&0.695\\
			$(rs1285057_{T2})$&0.521&1.029&0.507&0.613\\
			$(rs1437135_{G1})$&0.417&0.832&0.501&0.616\\
			$(rs1437135_{G2})$&1.338&1.491&0.897&0.370\\
			$(rs2544784_{T1})$&0.259&0.902&0.287&0.774\\
			$(rs2544784_{T2})$&1.544&1.248&1.237&0.216\\
			$(rs3765945_{G1})$&0.242&0.861&0.281&0.779\\
			$(rs3765945_{G2})$&0.828&1.051&0.788&0.431\\
			$rs439954_{A1}$&0.321&0.887&0.362&0.717\\
			$rs439954_{A2}$&1.808&1.213&1.491&0.136\\
			$(rs9322343_{G1})$&1.932&1.494&1.293&0.196\\
			$rs9371236_{G1}$&-1.119&2.143&-0.522&0.602\\
			$(rs9944225_{A1})$&0.661&0.895&0.739&0.460\\
			$(rs9944225_{A2})$&2.563&1.813&1.414&0.157\\
			$r(\mbox{rs1389930}_{C})$&3.303&2.087&1.583&0.113\\
			$r(\mbox{rs2013562}_{T})$&4.395&1.598&2.751& \textbf{0.00594}\\\hline

			\hline
		\end{tabular}
	\end{table}
	
	\begin{table}[!htbp]
		\centering
		
		\begin{tabular}{|c|c|c|c|c|}
			\hline
			\multicolumn{5}{|c|}{\texttt{Table 11 continued}} \\ \hline
			Coefficient & Estimate & Std.Error& Z value& Pr($>|z|$)\\\hline
			\multicolumn{5}{|c|}{\textbf{Testing the missingness mechanism of $rs7706809_{T}$}}\\\hline
			Intercept&-5.098&1.320&-3.861& \textbf{0.0001}\\
			$(rs3020348_{C1})$&-3.974&2.167&-1.834&0.067\\
			$(rs3020348_{C2})$&-2.897&2.913&-0.994&0.320\\
			$rs4646312_{C1}$&0.025&1.365&0.018&0.986\\
			$rs4646312_{C2}$&0.453&1.599&0.283&0.777\\     
			$rs4952220_{C1}$&3.628&1.228&2.955& \textbf{0.003}\\
			$rs4952220_{C2}$&1.718&2.107&0.815&0.415\\
			$(rs7085249_{G1})$&0.931&0.706&1.319&0.187\\  
			$(rs7085249_{G2})$&-0.680&1.390&-0.489&0.625\\   
			$(rs737866_{C1})$&0.970&1.007&0.964&0.335\\
			$(rs737866_{C2})$&1.310&1.509&0.868&0.385\\
			$rs936307_{A1}$&-0.713&1.312&-0.544&0.587\\
			$rs936307_{A2}$&0.568&2.001&0.284&0.776\\
			$r(\mbox{rs10916}_{C})$&2.941&1.505&1.953&0.051\\
			$r(\mbox{rs3775775}_{G})$&3.520&1.160&3.034& \textbf{0.002}\\
			$(rs165722_{C1})$&-0.512&4.122&0.124&0.901\\
			$(rs165722_{C2})$&-3.484&3.305&-1.054&0.292\\
			$(rs174675_{T1})$&0.110&0.938&0.117&0.907\\
			$(rs174675_{T2})$&3.005&1.228&2.448& \textbf{0.014}\\ 
			$rs2008691_{G1}$&0.811&2.033&0.399&0.690\\   
			$rs2008691_{G2}$&3.814&3.071&1.242&0.214\\
			$(rs2268796_{A1})$&-4.920&1.362&-3.611& \textbf{0.0003}\\
			$(rs2268796_{A2})$&-2.229&1.891&-1.179&0.239\\      
			$(rs2470176_{G1})$&-0.413&2.014&-0.205&0.837\\    
			$(rs2470176_{G2})$&1.400&3.267&0.428&0.668\\    
			$(rs2668854_{G1})$&2.104&0.843&2.497& \textbf{0.012}\\  
			$(rs2668854_{G2})$&4.182&2.019&2.071& \textbf{0.038}\\
			$rs3020343_{C1}$&3.393&2.157&1.573&0.116\\     
			$rs3020343_{C2}$&2.756&3.037&0.907&0.364\\
			$rs3797179_{A1}$&-0.478&0.784&-0.609&0.542\\  
			$rs3797179_{A2}$&2.521&1.245&2.025& \textbf{0.043}\\
			$(rs4680_{G1})$&-1.626&4.039&-0.397&0.691\\  
			$(rs4680_{G2})$&3.261&3.198&1.020& 0.308\\
			$(rs9371554_{C1})$&1.500&1.007&1.489& 0.137\\   
			$(rs9371554_{C2})$&-2.193&2.828&-0.775&0.438\\ 
			\hline                   
		\end{tabular}
	\end{table}

	\begin{table}[!htbp]
		\centering
		
		\begin{tabular}{|c|c|c|c|c|}
			\hline
            \multicolumn{5}{|c|}{\texttt{Table 11 continued}} \\ \hline
			Coefficient & Estimate & Std.Error& Z value& Pr($>|z|$)\\\hline
     \multicolumn{5}{|c|}{\textbf{Testing the missingness mechanism of $rs4702374_{G}$}}\\\hline  
			Intercept &-6.543&1.140&-5.740& \textbf{9.45e-9}\\
			$r(\mbox{rs1651074}_{A})$&4.328&1.598&2.708& \textbf{0.007}\\
			$r(\mbox{rs7706809}_T)$&1.422&1.603&0.887& 0.375\\
			$(rs117497814_{G1})$&0.902&0.713&1.266&0.205\\
			$(rs117497814_{G2})$&2.483&1.121&2.214& \textbf{0.027}\\		   
			$rs2255192_{T1}$&-0.654&0.786&-0.833&0.405\\
			$rs2255192_{T2}$&2.222&0.987&2.252& \textbf{0.024}\\
			$(rs2268796_{A1})$&-3.176&1.871&-1.698& 0.090\\
			$(rs2268796_{A2})$&-4.161&2.533&-1.643&0.100\\
			$rs2305707_{G1}$&2.392&1.997&1.198&0.231\\		
			$rs2305707_{G2}$&6.160&2.995&2.056& \textbf{0.040}\\
			$(rs2665584_{G1})$&2.052&0.738&2.779& \textbf{0.005}\\	
			$(rs2665584_{G2})$&3.824&1.878&2.036& \textbf{0.042}\\
			$rs4952220_{C1}$&2.185&1.428&1.529&0.126\\
			$rs4952220_{C2}$&9.910&2.396&2.468& \textbf{0.014}\\\hline
			\multicolumn{5}{|c|}{\textbf{Testing the missingness mechanism of $rs248805_{A}$}}\\\hline
			Intercept &-5.952&1.205&-4.937& \textbf{7.92e-7}\\
			$r(\mbox{rs1651074}_{A})$&5.958&2.125&2.803& \textbf{0.005}\\
			$r(\mbox{rs4702374}_{G})$&3.026&1.593&1.900& 0.057\\
			$r(\mbox{rs4702379}_{T})$&5.984&1.531&3.908& \textbf{9.31e-5}\\  
			$r(\mbox{rs7706809}_{T})$&-1.104&2.667&-0.414& 0.679\\       				$(rs10514523_{A1})$&-0.631&1.503&-0.433&0.665\\
			$(rs10514523_{A2})$&0.913&1.379&0.662&0.508\\	
			$rs2255192_{T1}$&-1.897&1.433&-1.324&0.185\\
			$rs2255192_{T2}$&3.135&1.296&2.419& \textbf{0.016}\\\hline
			\multicolumn{5}{|c|}{\textbf{Testing the missingness mechanism of $rs2758331_{A}$}}\\\hline
			Intercept &-7.061&1.640&-4.307& \textbf{1.65e-5}\\
			$(rs1112005_{T1})$&-0.509&1.072&-0.475&0.635\\
			$(rs1112005_{T2})$&0.442&1.173&0.377&0.706\\
			$(rs1651074_{A1})$&3.197&1.645&1.944&0.052\\
			$(rs1651074_{A2})$&5.401&2.613&2.067& \textbf{0.039}\\
			$(rs248800_{A1})$&-1.492&1.550&-0.963&0.336\\
			$(rs248800_{A2})$&-1.251&2.454&-0.510&0.610\\
			$(rs3765945_{G1})$&-0.344&1.177&-0.293&0.770\\
			$(rs3765945_{G2})$&1.956&1.030&1.899&0.058\\
			$(rs4702379_{T1})$&1.342&1.324&1.013&0.311\\ 
			$(rs4702379_{T2})$&4.014&1.594&2.518& \textbf{0.012}\\ 
			\hline                  
		\end{tabular}
	\end{table}

	\begin{table}[!htbp]
		\centering
		
		\begin{tabular}{|c|c|c|c|c|}
			\hline
			\multicolumn{5}{|c|}{\texttt{Table 11 continued}} \\ \hline
			Coefficient & Estimate & Std.Error& Z value& Pr($>|z|$)\\\hline
			\multicolumn{5}{|c|}{\textbf{Testing the missingness mechanism of $rs6163_{A}$}}\\\hline
			Intercept &-2.388&1.498&-4.931& \textbf{8.2e-7}\\
			$r(\mbox{rs4919686}_{C})$ &5.033&1.548&3.252& \textbf{0.001}\\
			$r(\mbox{rs7753153}_{A})$&4.455&1.983&2.246& \textbf{0.025}\\
			$r(\mbox{rs806645}_{T})$ &4.160&1.809&2.300& \textbf{0.021}\\
			$rs2255192_{T1}$&0.016&1.466&0.011& \textbf{0.001}\\ 
			$rs2255192_{T2}$&3.484&1.797&1.939&0.053 \\ \hline
			\multicolumn{5}{|c|}{\textbf{Testing the missingness mechanism of $rs4147581_{G}$}}\\\hline
			Intercept &-2.694&0.479&-5.621& \textbf{1.9e-8}\\
			$(rs2547231_{C1})$&-0.315&0.479&-0.658& 0.511\\
			$(rs2547231_{C2})$&1.849&0.826&2.239& \textbf{0.025}\\
			$(rs2665584_{G1})$&1.001&0.479&2.091& \textbf{0.037}\\
			$(rs2665584_{G2})$&2.370&1.248&1.898& 0.058\\
			$(rs2758331_{A1})$&-0.902&0.443&-2.037& \textbf{0.042}\\ 
			$(rs2758331_{A2})$&-1.092&0.540&-2.023& \textbf{0.043}\\ 
			$(rs408876_{A1})$&-0.733&0.528&-1.388& 0.165\\
			$(rs408876_{A2})$&1.465&0.755&1.942&0.052\\
			$(rs727479_{C1})$&-0.913&0.447&-2.044& \textbf{0.041}\\ 
			$(rs727479_{C2})$&-0.221&0.666&-0.331&0.740\\ 
			$(rs8041933_{A1})$&1.441&0.439&3.283& \textbf{0.001}\\
			$(rs8041933_{A2})$&-0.158&1.482&-0.107&0.915\\	
			\hline 
			\multicolumn{5}{|c|}{\textbf{Testing the missingness mechanism of $rs1004984_{A}$}}\\\hline
			Intercept &-6.864&1.203&-5.704& \textbf{1.17e-8}\\
			$r(\mbox{rs248800}_{A})$&2.592&1.857&1.395&0.163\\
			$r(\mbox{rs248805}_{A})$&2.291&2.434&0.941&0.347\\
			$r(\mbox{rs4646247}_{A})$ &6.6614&2.060&3.234& \textbf{0.001}\\
			$r(\mbox{rs4702379}_{T})$&1.4586&2.091&0.698&0.485\\
			$r(\mbox{rs6601927}_{G})$&0.404&1.754&0.230&0.818\\\hline
			\multicolumn{5}{|c|}{\textbf{Testing the missingness mechanism of $rs12917295_{C}$}}\\\hline
			Intercept &-9.833&2.377&-4.137& \textbf{3.52e-5}\\
			$r(\mbox{rs1134095}_{G})$ &5.240&1.802&2.908& \textbf{0.004}\\
			$r(\mbox{rs4147581}_{G})$&2.643&1.146&2.306& \textbf{0.021}\\
			$r(\mbox{rs6410}_{T})$&3.317&1.245&2.664& \textbf{0.008}\\
			$(rs1799929_{T1})$&2.276&1.954&1.165&0.244\\
			$(rs1799929_{T2})$&2.817&1.047&1.376&0.169\\
			$rs1983828_{G1}$&1.203&1.434&0.839&0.402\\
			$rs1983828_{G2}$&3.351&1.602&2.092& \textbf{0.036}\\
			$rs2476923_{A1}$&-0.161&1.419&-0.113&0.910\\
			$rs2476923_{A2}$&1.785&1.341&1.330&0.183\\
			\hline
		\end{tabular}
	\end{table}

\section{Discussion} \label{sec-5}
High dimensionality, high correlations, and widespread missing values with non-ignorable missingness mechanisms
are the three challenges present in GWAS. The main contribution of this paper is the development of
a coherent statistical procedure of categorical phenotype-genotype association analysis, integrating the
state-of-the-art methods of random forest for variable selection, weighted ridge regression with EM
algorithm for missing data imputation, and linear statistical hypothesis testing for determining the missingness
mechanism. Two simulated GWASs have been carried out to assess the performance of the proposed procedure, followed by
a real data analysis on breast cancer GWAS for illustration.

The statistical methods used to develop our 
\textit{Ridge-EM $\rightarrow$ multiple imputations $\rightarrow$ complete data $\rightarrow$ RF
$\rightarrow$ variable selection $\rightarrow$ Ridge-EM} procedure are well established and implemented
into several statistical computing environments. But integrating them to tackle GWAS data analysis involving
missing values with non-ignorable missingness mechanism is not trivial but requires dedicated efforts.
For example, the GeneSrF method by \citeA{diaz2006gene} is not able to handle non-uniform high-dimensional
imputation weights that are commonly present in missing data analysis. We have overcome this difficulty
by applying the fast algorithm of \citeA{wright2015ranger} to implement GeneSrF, and have obtained 
satisfactory outcomes. As another example, multiple imputation techniques have been used in GWAS to 
replace the missing values by surrogates; but the existing practice stops going further to investigate
whether the missingness mechanism is ignorable or not. We are able to test whether or not the 
missingness mechanism is ignorable by applying a standard statistical linear hypothesis testing procedure. 
This is only possible after we establish the model system (\ref{eq-1}, \ref{eq-4}, \ref{eq-8}) for GWAS.

It has been observed that our developed method would become computationally very intensive if the 
proportions of missing values in relevant SNPs are very high and number of such SNPs is also high, 
because of the combinatorial explosion involved in multiple imputation and variable selection. 
A promising solution to this difficulty is to incorporate Markov chain Monte Carlo into the EM
algorithm, and to perform variable selection by stochastic search. We are currently 
working on this solution, and will report the results somewhere else once available.

\bibliographystyle{apacite}
\bibliography{bibliography}
\end{document}